\documentclass[12pt]{iopart}
\usepackage{amssymb}
\usepackage{epsfig}
\usepackage{graphicx}
\begin{document}
\title[Coding of nonlinear states for the Gross-Pitaevskii equation]{Coding of nonlinear states for the Gross-Pitaevskii
equation with periodic potential}

\author{G L Alfimov, A I Avramenko}

\address{National Research University of Electronic Technology, Moscow, 124498, Russia}
\ead{galfimov@yahoo.com}

\begin{abstract}
We study nonlinear states for NLS-type equation with additional
periodic potential $U(x)$ (called {\it the Gross-Pitaevskii equation,
GPE} in theory of Bose-Einstein Condensate, BEC). We prove that if the
nonlinearity is defocusing (repulsive, in BEC context) then under
certain conditions there exists a homeomorphism between the set of
nonlinear states for GPE (i.e. real bounded solutions of some nonlinear
ODE) and the set of bi-infinite sequences of numbers from 1 to $N$ for
some integer $N$. These sequences can be viewed as codes of the
nonlinear states. Sufficient conditions for the homeomorphism to exist
are given in the form of three hypotheses. For a given $U(x)$, the
verification of the hypotheses should be done numerically. We report on
numerical results for the case of GPE with cosine potential and
describe regions in the plane of parameters where this coding is
possible.
\end{abstract}
\pacs{03.75.Lm, 05.45.-a, 02.30.Hq} \ams{37B10,35Q55,65P99}
\submitto{\NL} \maketitle

\maketitle

\section{Introduction}
The nonlinear Schr\"{o}dinger equation with additional potential
$U({\bf x})$,
\begin{eqnarray}
        &&i \Psi_t=-\Delta\Psi+U({\bf x})\Psi+\sigma |\Psi|^2\Psi,\label{GPEq}\\
        &&\Delta\equiv\frac{\partial^2\,}{\partial
x^2}+\frac{\partial^2\,}{\partial
        y^2}+\frac{\partial^2\,}{\partial z^2},\quad \sigma=\pm 1
\nonumber
\end{eqnarray}
arises in many physical applications including models of optics
\cite{Optics}, plasma physics \cite{Berge} and  theory of ultracold
gases \cite{PitStr}. In the last context, Eq.(\ref{GPEq}) (called {\it
the Gross-Pitaevskii equation}, GPE) appears as one of the basic
equations to describe the phenomenon of Bose-Einstein condensation
(BEC) in so-called mean-field approximation. In this case $\Psi(t, {\bf
x})$ means the macroscopic wave function of the condensate, $\sigma=1$
corresponds to the case of to repulsive interparticle interactions and
$\sigma=-1$ - to the case of attractive interactions. The function
$U({\bf x})$ describes the potential of the trap to confine the
condensate.  In particular, magnetic trap has been modelled by the
parabolic potential $U({\bf x})=|{\bf x}|^2$ and optical trap has been
described by the potential which is periodic with respect to one or
several variables \cite{KonotSurvey,PitUsp,MorshOb}.

An important class of solutions for Eq.(\ref{GPEq}) are stationary
nonlinear states defined by the anzats
\begin{equation}
\Psi(t,{\bf x})=e^{-i\omega t}\psi({\bf x}).\label{Anzats}
\end{equation}
The parameter $\omega$ in terms of BEC corresponds to the chemical
potential. The function $\psi({\bf x})$ solves the equation
\begin{equation}
\Delta\psi+(\omega-U({\bf x}))\psi-\sigma |\psi|^2\psi
=0.\label{StatGPEq}
\end{equation}
It is known that Eq.(\ref{StatGPEq}) describes a great variety of
nonlinear objects. In particular, it has been found that real
1D-version of Eq.(\ref{StatGPEq})
\begin{eqnarray}
\psi_{xx}+(\omega-U(x))\psi-\sigma\psi^3=0 \label{1D}
\end{eqnarray}
with the model cosine potential
\begin{eqnarray}
U(x)=A\cos 2x \label{cos-pot}
\end{eqnarray}
 describes bright and dark gap solitons
\cite{Kiv2003,Dark,AKS,PelKiv}, nonlinear periodic structures
(nonlinear Bloch waves) \cite{Kiv2003,BlochW}, domain walls
\cite{DomWalls05}, gap waves \cite{Kiv2006} and so on.  Some
interesting relations between various nonlinear objects described by
Eq.(\ref{1D}) have been observed. In particular, in papers
\cite{Zhang09_1,Zhang09_2} the {\it composition relation} between gap
solitons and nonlinear Bloch waves was established: it has been
observed that a nonlinear Bloch wave can be approximated by an infinite
chain of narrow gap solitons (called {\it fundamental gap solitons},
FGS), each localized in one well of the periodic potential. In
\cite{China03_2011} this principle has been applied to the case of more
general nonlinearity. It is worth noting that the results of
\cite{Kiv2006} also can be interpreted in a similar sense, since the
gap waves discovered in \cite{Kiv2006} can be regarded as compositions
of finite number of FGS.


In the present paper, we address the problem of  description of
nonlinear states covered by Eq.(\ref{1D}) in the case of repulsive
interactions, $\sigma=1$, i.e for the equation
\begin{eqnarray}
\psi_{xx}+(\omega-U(x))\psi-\psi^3=0. \label{1D_rep}
\end{eqnarray}
We argue that if the periodic potential $U(x)$ satisfies some
conditions, then {\it all} the solutions of Eq.(\ref{1D_rep}) defined
at the whole $\mathbb{R}$ can be put in one-to-one correspondence with
bi-infinite sequences of integers $n=1,\ldots,N$ (called {\it codes}).
The correspondence is a homeomorphism for properly introduced
topological spaces. Each of the integers $n$ is ``responsible'' for the
behavior of the solution $\psi(x)$ on one period of the potential
$U(x)$. From this viewpoint, the solutions $\psi(x)$ may be regarded as
compositions of FGS localized in the wells of the periodic potential
and taken with a proper sign. So, the coding technique gives a unified
approach to describe both gap solitons and nonlinear Bloch waves and
generalizes (and justifies) the composition relation of
\cite{Zhang09_1,Zhang09_2}. In order to conclude that for a given
$U(x)$ the coding is possible, one has to verify numerically three
Hypotheses formulated in Section \ref{CodingGen}. As an example, we
applied this method to the case of model periodic potential
(\ref{cos-pot}) and present the regions in the parameter plane
$(\omega,A)$ where all nonlinear states can be encoded with bi-infinite
sequences of integers.

Our approach is based on the following observation: the ``most part''
of the solutions for Eq.(\ref{1D_rep}) are {\it singular}, i.e. they
collapse (tend to infinity) at some finite point of real axis. The set
of initial data at $x=0$ for non-collapsing solutions  can be found
numerically by properly organized scanning procedure. Then we study
transformations of this set under the action of Poincare map using
methods of symbolic dynamics. A similar idea was used to justify a
strategy of ``demonstrative computations'' of nonlinear modes for 1D
GPE with repulsive interactions and multi-well potential
\cite{AlfZez,AlfZez1}. This allowed to find numerically all the
localized modes for Eq.(\ref{1D_rep}) with single-well and double-well
potentials and to guarantee that no other localized modes exist.

The paper is organized as follows. In Section \ref{singular} we
introduce some notations and definitions which will be used throughout
the rest of the text and make some assertions about them. In Section
\ref{SymDynTheory} we formulate a theorem (Theorem
\ref{SymDynTheory}.1) which gives a base for our method. Section
\ref{CodingGen} contains an application of Theorem \ref{SymDynTheory}.1
to the case of Eq.(\ref{1D_rep}). The main outcome of Section
\ref{CodingGen} is formulated in the form of three hypotheses. These
hypotheses provide sufficient conditions for the homeomorphism
mentioned above to exist and should be verified numerically. In Section
\ref{Cosine} we apply this approach to the case of the cosine potential
(\ref{cos-pot}). Section \ref{concl} includes summary and discussion.

For the sake of clarity all the proofs are removed from the main text
to Appendices.

\section{Bounded and singular solutions}\label{singular}

\subsection{Some definitions}\label{Def}

In what follows we refer to a solution $\psi(x)$ of Eq.(\ref{1D_rep})
as {\it a singular solution} if for some $x=x_0$
\begin{eqnarray*}
\lim_{x\to x_0}\psi(x)=+\infty\quad \mbox{or}\quad \lim_{x\to
x_0}\psi(x)=-\infty.
\end{eqnarray*}
In this case we say that the solution $\psi(x)$ {\it collapses} at
$x_0$. Also let us introduce the following definitions:
\medskip

\medskip

{\it Collapsing and non-collapsing points:}  A point $(\psi_0,\psi_0')$
of the plane $\mathbb{R}^2=(\psi,\psi')$ is
\begin{itemize}
\item  {\sl  $L$-collapsing forward
point}, $L>0$, if the solution of Cauchy problem for Eq.(\ref{1D_rep})
with initial data $\psi(0)=\psi_0$, $\psi_x(0)=\psi_0'$ collapses at
value $x=x_0$ and $0<x_0< L$;
\item {\sl  $L$-non-collapsing forward point}, $L>0$, if
the solution of Cauchy problem for Eq.(\ref{1D_rep}) with initial data
$\psi(0)=\psi_0$, $\psi_x(0)=\psi_0'$ does not collapse at any value
$x=x_0$, $0<x_0\leq L$.
\item
{\sl $L$-collapsing backward point} if the corresponding solution of
Cauchy problem for Eq.(\ref{1D_rep}) collapses  at some value $x=-x_0$
and $0<x_0< L$;
\item
{\sl $L$-non-collapsing backward point} if the corresponding solution
of Cauchy problem for Eq.(\ref{1D_rep}) does not collapse  at any value
$x=-x_0$, $0<x_0\leq L$;
\item {\sl $\infty$-non-collapsing forward/backward point} if it is not $L$-collapsing forward/backward point for any $L>0$;
\item {\sl $\infty$-non-collapsing point} if it is a $\infty$-non-collapsing
forward and backward point simultaneously;
\item
{\sl a collapsing point} if it is either $L$-collapsing forward or
backward for some $L$.
\end{itemize}

{\it Functions $h^\pm(\tilde{\psi},\tilde{\psi}')$.} The functions
$h^+(\tilde{\psi},\tilde{\psi}')$ and $h^-(\tilde{\psi},\tilde{\psi}')$
are defined in ${\mathbb{R}}^2$ as follows:
$h^+(\tilde{\psi},\tilde{\psi}')=x_0$ if the solution of Cauchy problem
for Eq.(\ref{1D_rep}) with initial data $\psi(0)=\tilde{\psi}$,
$\psi_x(0)=\tilde{\psi}'$ collapses at value $x=x_0$, $x_0>0$. By
convention, we assume that $h^+(\tilde{\psi},\tilde{\psi}')=\infty$ if
$(\tilde{\psi},\tilde{\psi}')$ is $\infty$-non-collapsing forward
point. Similarly, $h^-(\tilde{\psi},\tilde{\psi}')=-x_0$ if the
solution of Cauchy problem for Eq.(\ref{1D_rep}) with initial data
$\psi(0)=\tilde{\psi}$, $\psi_x(0)=\tilde{\psi}'$ collapses at value
$x=x_0$, $x_0<0$.\medskip

{\it The sets ${\cal U}_L^\pm$ and ${\cal U}_L$.} We denote the set of
all $L$-non-collapsing forward points by ${\cal U}_L^+$ and the set of
all $L$-non-collapsing backward points by ${\cal U}_L^-$. In terms of
the functions $h^\pm(\psi,\psi')$ these sets are
$$
{\cal U}^+_L=\{(\psi,\psi')\in\mathbb{R}^2|~ h^+(\psi,\psi')> L\},\quad
{\cal U}^-_L=\{(\psi,\psi')\in\mathbb{R}^2|~ h^-(\psi,\psi')> L\}.
$$
The intersection of ${\cal U}_L^+$ and ${\cal U}_L^-$ will be denoted
by ${\cal U}_L$. Evidently, if $L_1<L_2$ then ${\cal U}^+_{L_2}\subset
{\cal U}^+_{L_1}$, ${\cal U}^-_{L_2}\subset {\cal U}^-_{L_1}$ and
${\cal U}_{L_2}\subset {\cal U}_{L_1}$.\medskip

{\it The values $\overline{\Omega}$ and $\underline{\Omega}$.} We
define
\begin{eqnarray*}
\overline{\Omega}\equiv\sup_{x\in{\mathbb{R}}}(\omega-U(x)),\quad
\underline{\Omega}\equiv\inf_{x\in{\mathbb{R}}}(\omega-U(x)).
\end{eqnarray*}

\subsection{Some statements about collapsing
points}\label{StatNonCol}

In what follows we will use some statements from the paper
\cite{AlfZez}, in particular so-called Comparison Lemma (reproduced in
\ref{Th1} for convenience). It is known \cite{AlfZez} that for
$\overline{\Omega}<0$ Eq.(\ref{1D_rep}) has no bounded on $\mathbb{R}$
solutions, therefore we restrict our analysis by the case
$\overline{\Omega}\geq0$. Also it is known (\cite{AlfZez}, Lemma 2)
that all $\infty$-non-collapsing points for Eq.(\ref{1D_rep}) are
situated in the strip $-\sqrt{\overline{\Omega}}\leq\psi\leq
\sqrt{\overline{\Omega}}$. Theorem \ref{singular}.1 below gives more
detailed information about collapsing points for Eq.(\ref{1D_rep}).
\medskip

{\bf Theorem \ref{singular}.1.} {\it  Let the potential $U(x)$ be
continuous and bounded on $\mathbb{R}$. Then for each $L$ there exist
$\tilde{\psi}_L$ and $\tilde{\psi}'_L$ such that the set ${\cal U}_L$
is situated in the rectangle $-\tilde{\psi}_L<\psi<\tilde{\psi}_L$,
$-\tilde{\psi}'_L<\psi'<\tilde{\psi}'_L$.}
\medskip

The proof of Theorem \ref{singular}.1 is quite technical. We postponed
it in \ref{Th1}.

Another important statement is as follows:
\medskip

{\bf Theorem \ref{singular}.2.} {\it Let the potential $U(x)$ be
continuous and bounded on $\mathbb{R}$ and
$h^+(\psi_0,\psi'_0)=L<\infty$. Then $h^+(\psi,\psi')$ is a continuous
function in some vicinity of the point $(\psi_0,\psi'_0)$.}
\medskip

The proof of Theorem \ref{singular}.2 can be found in \ref{Th2}. It is
worth commenting Theorem \ref{singular}.2 as follows:\medskip

(i) Analogous statement is valid for the function
$h^-(\psi,\psi')$.\medskip

(ii) It follows from Theorem \ref{singular}.2 that if the potential
$U(x)$ is continuous and bounded on $\mathbb{R}$ then ${\cal U}_L^\pm$
and ${\cal U}_L$ are open sets. The boundary of the set ${\cal U}_L^+$
consists of continuous curves and corresponds to the level lines
$h^+(\psi,\psi')=L$ of the function $h^+(\psi,\psi')$. This boundary
consists of the points $(\tilde{\psi},\tilde{\psi}')$ such that the
solution of Eq.(\ref{1D_rep}) with initial data $\psi(0)=\tilde{\psi}$,
$\psi_x(0)=\tilde{\psi}'$ satisfies one of the conditions
\begin{eqnarray*}
\lim_{x\to L}\psi(x)=+\infty\quad \mbox{or}\quad \lim_{x\to
L}\psi(x)=-\infty.
\end{eqnarray*}
Correspondingly, the boundary of the set ${\cal U}_L^-$ is also
continuous and consists of the points $(\tilde{\psi},\tilde{\psi}')$
such that the solution of Eq.(\ref{1D_rep}) with initial data
$\psi(0)=\tilde{\psi}$, $\psi_x(0)=\tilde{\psi}'$ satisfies similar
conditions
\begin{eqnarray*}
\lim_{x\to -L}\psi(x)=+\infty\quad \mbox{or}\quad \lim_{x\to
-L}\psi(x)=-\infty.
\end{eqnarray*}

(iii) Theorem \ref{singular}.2 does not impose any restriction to the
behavior of $h^+(\psi,\psi')$ in a vicinity of a point
$(\psi_0,\psi'_0)$ where $h^+(\psi_0,\psi'_0)=\infty$. In practice,
this behavior may be very complex, see Sect.\ref{npi-noncol}.\medskip

The set of solutions for Eq.(\ref{1D_rep}) that collapse at a given
point $x=x_0$ can be described more precisely in terms of asymptotic
expansions.
\medskip

\begin{figure}
\centerline{\includegraphics [scale=0.7]{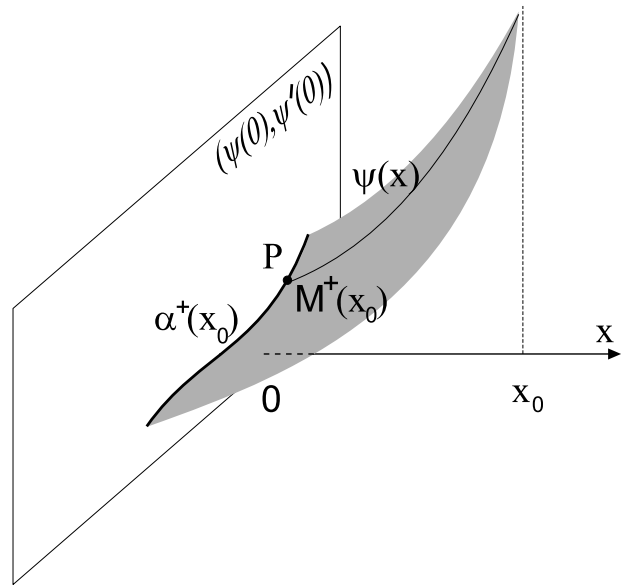}} \caption{The
point $P$, the manifold $M^+(x_0)$, the curve
$\alpha^+(x_0)$.}\label{CloseSet}
\end{figure}

{\bf Theorem \ref{singular}.3.} {\it Let  $x=x_0$ be an arbitrary fixed
real. Assume that $\tilde{U}(x)=\omega-U(x)$ in a vicinity of $x=x_0$
can be represented as follows
\begin{eqnarray*}
\tilde{U}(x)=U_0+U_1\delta +U_2 \delta^2+U_3 \delta^3+o\left(\delta
^3\right)
\end{eqnarray*}
where $\delta\equiv x-x_0$. Then the solutions of Eq.(\ref{1D_rep})
which satisfy the condition
\begin{eqnarray}
\lim_{x\to x_0}\psi(x)=+\infty \label{+infty}
\end{eqnarray}
obey the asymptotic expansion
\begin{eqnarray}
\psi(\delta)&=&\frac{\sqrt{2}}{\delta}+A_1\delta+A_2\delta
^2+A_3\delta^3\ln |\delta|+ {\bf C}\delta^3+ A_4 \delta^4+o\left(\delta
^4\right).\label{ExpanArb}
\end{eqnarray}
Here ${\bf C}\in\mathbb{R}$ is a free parameter and
\begin{eqnarray*}
A_1=\frac{\sqrt{2}U_0}6;\quad A_2=\frac{\sqrt{2}U_1}4;\quad
A_3=-\frac{\sqrt{2}U_2}5,\quad
A_4=\frac{\sqrt{2}}6\left(\frac{U_0U_1}{12}-U_3\right).
\end{eqnarray*}
}

{\it Proof:} The result follows from straightforward substitution of
series (\ref{ExpanArb}) into Eq.(\ref{1D_rep}). $\blacksquare$

Theorem \ref{singular}.3 should be commented as follows.\medskip

(i) The free parameter  ${\bf C}$ is ``internal'' parameter of
continuous one-parameter set of solutions which tend to $+\infty$ at
the point $x=x_0$. This situation can be illustrated by the following
heuristic reasoning. Let $P=(\psi_0,\psi_0')$ be a collapsing point and
the solution $\psi(x)$ of Cauchy problem for Eq.(\ref{1D_rep}) with
initial data $\psi(0)=\psi_0$, $\psi_x(0)=\psi_0'$ collapses at
$x=x_0>0$. Then, generically $\psi(x)$ belongs to a continuous
one-parameter set of solutions which also satisfy the condition
(\ref{+infty}) and obey the expansion (\ref{ExpanArb}). In 3D space
$(x,\psi,\psi')$ this set generates 2D manifold $M^{+}(x_0)$ (see
Fig.\ref{CloseSet}). Intersection of $M^{+}(x_0)$ with the plane
$(\psi,\psi')$ at $x=0$ includes the point $P$ and is non-empty.
Generically, this intersection in some vicinity of $P$ is {\it an 1D
curve} which we denote $\alpha^+(x_0)$. In the plane $(\psi,\psi')$
this curve corresponds to the level line $h^+(\psi,\psi')=x_0$.
\medskip

(ii) Since Eq.(\ref{1D_rep}) is invariant with respect to the symmetry
$\psi\to-\psi$, the solutions of Eq.(\ref{1D_rep}) which satisfy the
condition
\begin{eqnarray*}
\lim_{x\to x_0}\psi(x)=-\infty
\end{eqnarray*}
obey the same, up to sign, asymptotic expansion (\ref{ExpanArb}). The
corresponding manifold $M^{-}(x_0)$ and the curve $\alpha^-(x_0)$ are
defined in the same way.\medskip

\subsection{Example: the sets ${\cal U}^\pm_L$ and ${\cal U}_L$ for the cosine
potential}\label{Ex:cos_pot}

\begin{figure}
\centerline{\includegraphics [scale=0.8]{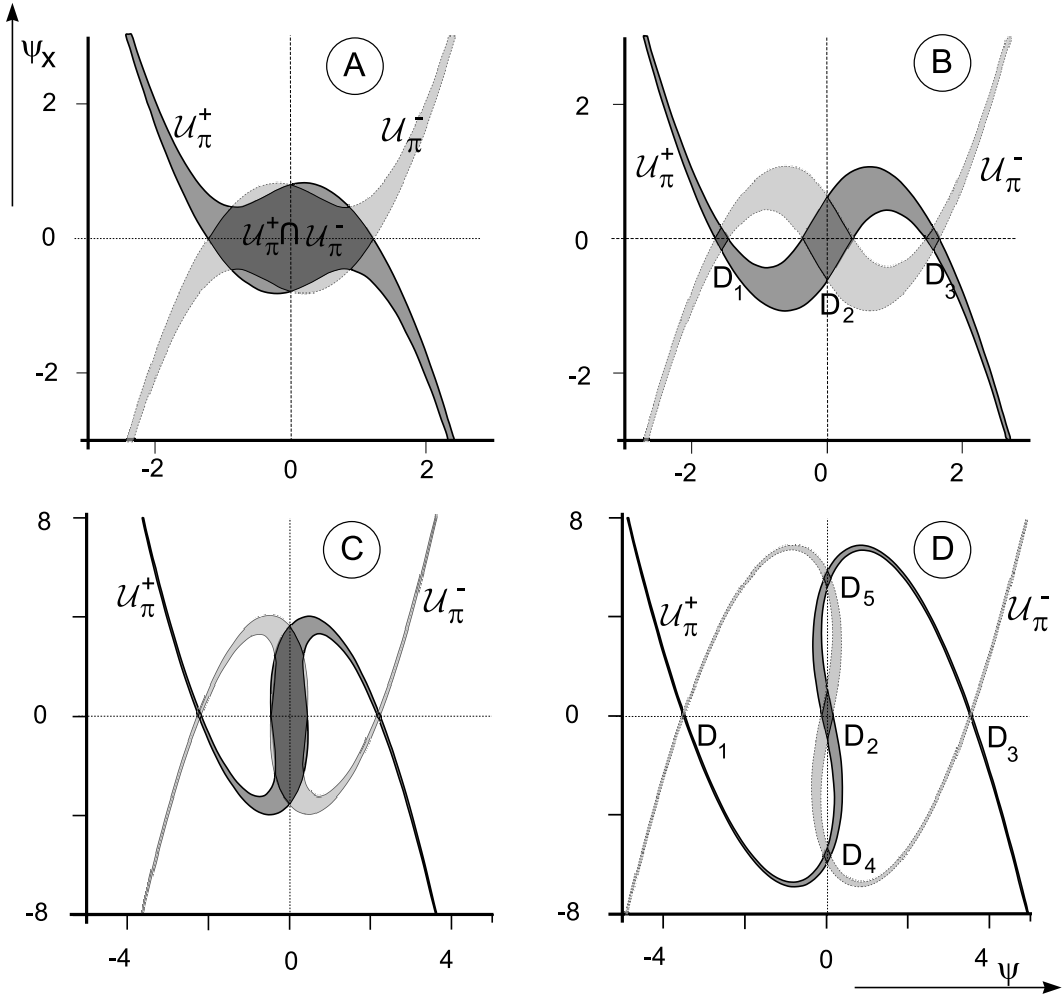}}\bigskip
\caption{The sets ${\cal U}_\pi^+$ and ${\cal U}_\pi^-$ for
Eq.(\ref{1Dcos}) and the parameters $\omega$ and $A$ lying in the first
(panels A and B) and the second (panels C and D) gaps, see
Sect.\ref{Cosine} for detail. The parameters are: (A) $\omega=1$,
$A=-1$; (B) $\omega=1$, $A=-3$; (C) $\omega=4$, $A=-4$; (D) $\omega=4$,
$A=-10$. The sets were obtained numerically by scanning of initial data
plane for Eq.(\ref{1Dcos}). The areas ${\cal U}_\pi={\cal U}_\pi^+\cap
{\cal U}_\pi^-$ are shown in dark.} \label{U_pi}
\end{figure}

Let us now give now examples of the sets ${\cal U}^\pm_L$ for
Eq.(\ref{1D_rep}) in the case of the cosine potential (\ref{cos-pot}).
Eq.(\ref{1D_rep}) takes the form
\begin{eqnarray}
\psi_{xx}+(\omega-A\cos2x)\psi-\psi^3=0. \label{1Dcos}
\end{eqnarray}
The sets ${\cal U}^\pm_L$ possess the following symmetry properties:
\begin{itemize}
\item [1.] Since the nonlinearity in Eq.(\ref{1Dcos}) is odd, both the sets ${\cal U}_L^\pm$
are symmetric in the plane $(\psi,\psi')$ with respect to
the origin;%
\item [2.] Since Eq.(\ref{1Dcos})  is invariant with respect to
$x$-inversion the sets ${\cal U}_L^+$ and ${\cal U}_L^-$ are related to
each other by the symmetry with respect to the axis $\psi$ and due to
p.1, with respect to the axis $\psi'$ also.
\end{itemize}

\subsubsection{$\pi$-non-collapsing forward/backward points of
Eq.(\ref{1Dcos}).}\label{pi-noncol}

\begin{figure}
\centerline{\includegraphics [scale=0.8]{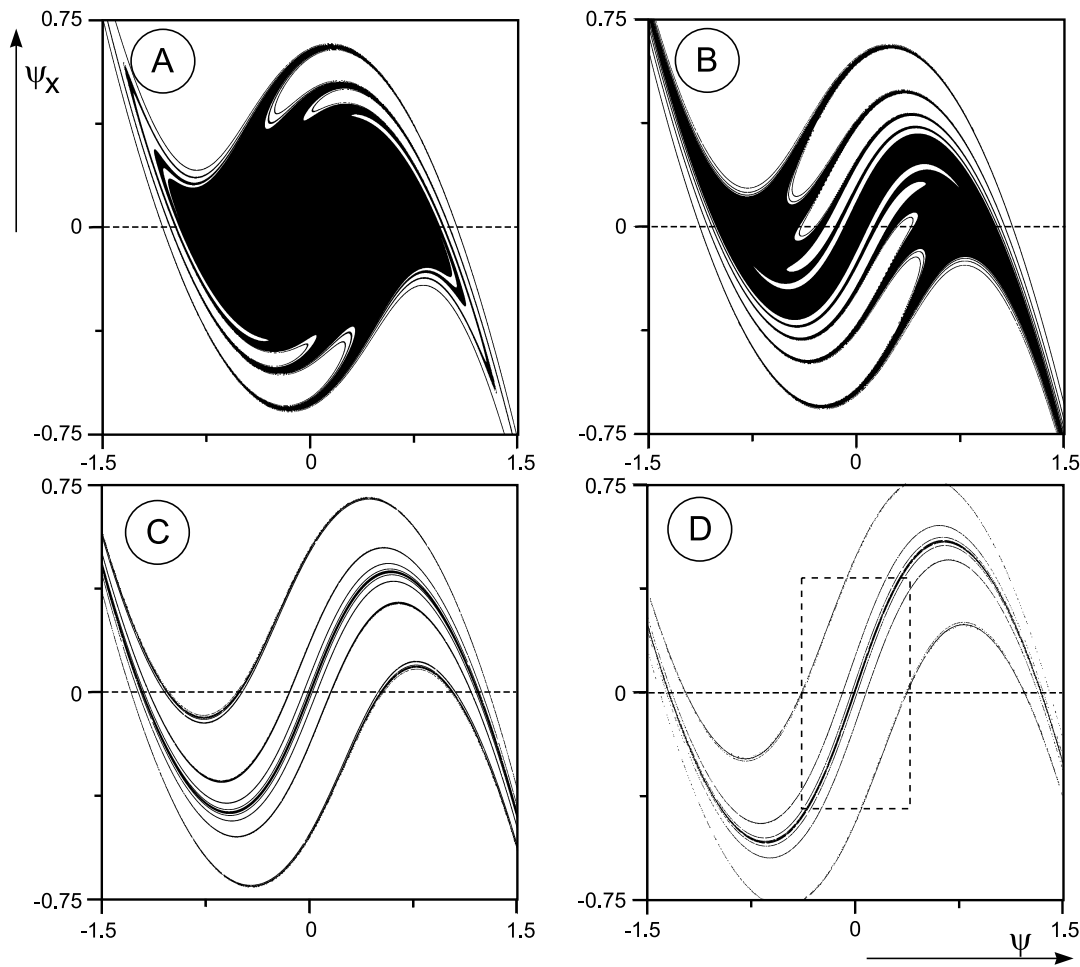}} \caption{The
sets ${\cal U}_{6\pi}^+$ for Eq.(\ref{1Dcos}). In all the cases
$\omega=1$ and (A) $A=-0.4$; (B) $A=-0.7$; (C) $A=-1.5$; (D) $A=-2.0$.
The rectangle in panel D is shown magnified in
Fig.\ref{Codes}.}\label{Fractals}
\end{figure}

The sets ${\cal U}_\pi^\pm$ were found by thorough numerical scanning
in the plane of initial data $(\psi,\psi')$ (some details of numerical
procedure can be found in Sect.\ref{Cosine}). The numerical study shows
that for any values of parameters $\omega$ and $A$ the sets ${\cal
U}_\pi^\pm$ are {\it infinite curvilinear strips}. The typical shapes
of the sets ${\cal U}_\pi^\pm$ for Eq.(\ref{1Dcos}) are shown in
Fig.\ref{U_pi}. The boundary of ${\cal U}_\pi^+$ is represented by two
continuous curves $\alpha^\pm$. The curve $\alpha^+$ consists of such
points $(\psi_0,\psi'_0)$ that the solution $\psi(x)$ of the Cauchy
problem for Eq.(\ref{1Dcos}) with initial data $\psi(0)=\psi_0$,
$\psi'(0)=\psi'_0$ collapses at $x=\pi$ and
$\lim_{x\to\pi}\psi(x)=+\infty$. At the curve $\alpha^-$ the solution
$\psi(x)$ of the corresponding Cauchy problem obeys the condition
$\lim_{x\to\pi}\psi(x)=-\infty$. Similarly, the boundary of ${\cal
U}_\pi^-$ is represented by two continuous curves $\beta^\pm$. The
curves $\beta^\pm$ consist of points $(\psi_0,\psi'_0)$ such that the
solution $\psi(x)$ of the Cauchy problem for Eq.(\ref{1Dcos}) with
initial data $\psi(0)=\psi_0$, $\psi'(0)=\psi'_0$ collapses at $x=-\pi$
and $\lim_{x\to-\pi}\psi(x)=\pm\infty$.

$\pi$-non-collapsing forward {\it and} backward points of
Eq.(\ref{1Dcos}) form the set ${\cal U}_\pi={\cal U}_\pi^+\cap{\cal
U}_\pi^-$. It follows from Fig.\ref{U_pi} that this set may consist of
several disjoined components. More detailed discussion of the sets
${\cal U}_{\pi}^\pm$  and ${\cal U}_{\pi}$ is postponed to
Sect.\ref{Cosine}.

\subsubsection{$\pi n$-non-collapsing forward/backward points of
Eq.(\ref{1Dcos}), $n>1$.}\label{npi-noncol}  Fig.\ref{Fractals}
exhibits the sets ${\cal U}_{6\pi}^+$ for $\omega=1$ and various values
of $A$. The sets ${\cal U}_{6\pi}^-$ are the reflections of the sets
${\cal U}_{6\pi}^+$ with respect to the $\psi$ axis. It follows from
Fig.\ref{Fractals} that the sets ${\cal U}_{6\pi}^\pm$ have quite a
complex layered structure. When $n$ grows, the structure of ${\cal
U}_{n\pi}^\pm$ becomes more complex resembling {\it fractals}. The
situation is similar to one described in \cite{DeltaComb} for
Eq.(\ref{1D}) in the case of delta-comb potential.


\section{Symbolic dynamics: theory}\label{SymDynTheory}

In this section we give a theoretical background  for description of
the non-collapsing solutions of Eq.(\ref{1D_rep}) in terms of symbolic
dynamics. Results of such kind are well-known in dynamical system
theory. The language and the technique go back to 60-70-ties, see e.g.
\cite{Moser,Alexeev,Wiggins}. In fact, the conditions which we
formulate (Theorem \ref{SymDynTheory}.1) can be regarded as some
version of the Conley-Moser conditions, see e.g. \cite{Wiggins}. A
peculiarity of the statement which we give below is that it is
convenient for direct numerical check in practice.

Let $(\psi,\psi')$ be Cartesian coordinates in $\mathbb{R}^2$ and
$\mu(S)$ be a measure of set $S$ in $\mathbb{R}^2$. Remind that a
function $f(x)$ is called {\it $\gamma$-Lipschitz function} if for any
$x_1$ and $x_2$ the relation holds
\begin{eqnarray*}
|f(x_2)-f(x_1)|\leq\gamma|x_2-x_1|.
\end{eqnarray*}
Also introduce the following definitions.\medskip

{\it Definition.} Let $\gamma$ be a fixed real. We call {\it an island}
an open curvilinear quadrangle $D\subset \mathbb{R}^2$ formed by
nonintersecting  curve segments $\alpha^+$, $\beta^+$, $\alpha^-$,
$\beta^-$ ($\alpha^+$ and $\alpha^-$ are opposite sides of the
quadrangle and have no common points as well as $\beta^+$ and
$\beta^-$) such that
\begin{itemize}
\item the segments $\alpha^+$ and $\alpha^-$ are graphs
of monotone non-decreasing/non-increasing $\gamma$-Lipschitz functions
$\psi'=v_\pm(\psi)$;
\item the segments $\beta^+$ and
$\beta^-$ are  graphs of monotone non-increasing/non-decreasing
$\gamma$-Lipschitz functions $\psi=w_\pm(\psi')$;
\item
if $v_\pm(\psi)$ are non-decreasing functions then $w_\pm(\psi')$ are
non-increasing functions and vice versa.
\end{itemize}
\medskip

{\it Definition.} Let $\gamma$ be a fixed real and $D$ be an island
bounded by curve segments $\alpha^+$, $\beta^+$, $\alpha^-$, $\beta^-$.
We call {\it v-curve} a segment of curve $\beta$ with endpoints on
$\alpha^-$ and $\alpha^+$ which
\begin{itemize}
\item is a graph of monotone non-decreasing/non-increasing
$\gamma$-Lipschitz function $\psi'=v(\psi)$;
\item if $\beta^\pm$ are graphs of monotone non-decreasing functions,
then $v(\psi)$ is also a monotone non-decreasing function. If
$\beta^\pm$ are graphs of monotone non-increasing functions, then
$v(\psi)$ is also a monotone non-increasing function.
\end{itemize}
Similarly, we call {\it h-curve} a segment of curve with endpoints on
$\beta^-$ and $\beta^+$ which
\begin{itemize}
\item is a graph of monotone non-increasing/non-decreasing
$\gamma$-Lipschitz function $\psi=w(\psi')$;
\item if $\alpha^\pm$ are graphs of monotone non-decreasing functions,
then $w(\psi')$ is also a monotone non-decreasing function. If
$\alpha^\pm$ are graphs of monotone non-increasing functions, then
$w(\psi')$ is also a monotone non-increasing function.
\end{itemize}

{\it Definition.}  Let  $D$ be an island. We call {\it v-strip} a
curvilinear strip contained between two nonintersecting v-curves,
including both v-curves. Similarly, we call {\it h-strip} an open
curvilinear strip contained between two nonintersecting h-curves,
including both h-curves.\medskip

\begin{figure}
\centerline{\includegraphics [scale=0.8]{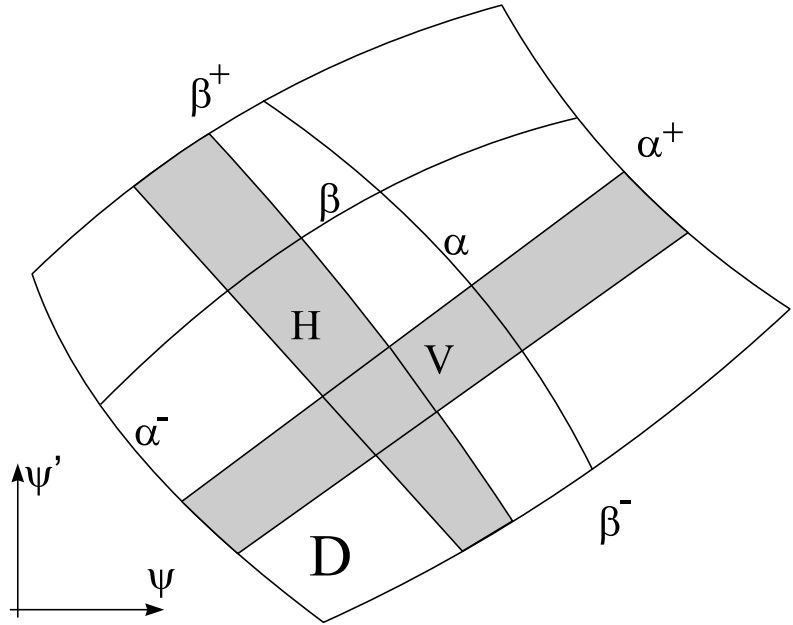}} \caption{An
island $D$ with v-curve $\beta$, v-strip $V$, h-curve $\alpha$ and
h-strip $H$. }\label{Island(def)}
\end{figure}

Fig.\ref{Island(def)} illustrates schematically the definitions
introduced above.

Let us denote $\Omega^N$ the set of bi-infinite sequences
$\{\ldots,i_{-1},i_0,i_1,\ldots\}$ where $i_k\in\{1,2,\ldots,N\}$.
$\Omega^N$ has the structure of topological space where the
neighborhood of a point $a^*=\{\ldots,i^*_{-1},i^*_0,i^*_1,\ldots\}$ is
defined by the sets
\begin{eqnarray*}
W_k(a^*)=\{a\in \Omega^N|~ i_j=i^*_j,|j|<k\}.
\end{eqnarray*}
Let  $T$ be a diffeomorphism  defined on a set $D=\bigcup_{i=1}^N D_i$
where each $D_i\subset \mathbb{R}^2$, $i=1,\ldots,N$, is an island and
all the islands are disjoined. Introduce the set $\cal P$ of
bi-infinite sequences (called {\it orbits})
\begin{eqnarray*}
{\bf s}=\{\ldots,p_{-1},p_0,p_1,\ldots\},\quad Tp_i=p_{i+1},
\end{eqnarray*}
where each $p_i=(\psi_i,\psi'_i)$, $i=0,\pm1,\pm 2,\ldots$, belongs to
$D$. $\cal P$ has the structure of metric space with norm defined as
\begin{eqnarray*}
\|{\bf s}\|=\sqrt{\psi_0^2+(\psi'_0)^2}.
\end{eqnarray*}
Define a map $\Sigma:{\cal P}\to \Omega^N$ as follows: $i_k$ is the
number $i$ of the component $D_i$ where the point $p_k$ lies. The
following statement is valid:\medskip

{\bf Theorem \ref{SymDynTheory}.1.} {\it Assume that:

(i) a diffeomorphism $T$ is defined on a set of $N$ disjoined islands
$D_i$, $i=1,\ldots,N$, $D=\bigcup_{i=1}^N D_i$; \medskip



(ii) for any $i$, $i=1,\ldots,N$, and for each v-strip $V\in D_i$ the
intersection $TV\cap D_j$, $j=1,\ldots,N$ is non-empty  and is also a
v-strip. Similarly, for any $i$, $i=1,\ldots,N$, and for each h-strip
$H\in D_i$ the intersection $T^{-1}H\cap D_j$, $j=1,\ldots,N$ is
non-empty  and is also an h-strip.\medskip

(iii) for the sequences of sets defined recurrently
\begin{eqnarray*}
&&\Delta^+_0=D,\quad \Delta^+_n=T\Delta^+_{n-1}\cap D,\\
&&\Delta^-_0=D,\quad \Delta^-_n=T^{-1}\Delta^-_{n-1}\cap D
\end{eqnarray*}
the conditions hold
\begin{eqnarray*}
\lim_{n\to\infty}\mu(\Delta^+_n)=0,\quad
\lim_{n\to\infty}\mu(\Delta^-_n)=0.
\end{eqnarray*}

Then $\Sigma$ is a homeomorphism between the topological spaces $\cal
P$ and $\Omega^N$.

}
\medskip

Proof of Theorem \ref{SymDynTheory}.1 is postponed in \ref{ProofMain}.


\section{Coding of solutions}\label{CodingGen}

\subsection{Poincare map}\label{Poincare}

Assume now that the potential $U(x)$ is {\it continuous} and {\it
$\pi$-periodic}
\begin{eqnarray*}
U(x+\pi)=U(x).
\end{eqnarray*}
{\it The Poincare map} $T:\mathbb{R}^2\to\mathbb{R}^2$ associated with
Eq.(\ref{1D_rep}) is defined as follows: if
$p=(\tilde\psi,\tilde\psi')\in \mathbb{R}^2$ then
$Tp=(\psi(\pi),\psi_x(\pi))$ where $\psi(x)$ is a solution of
Eq.(\ref{1D_rep}) with initial data $\psi(0)=\tilde\psi$,
$\psi_x(0)=\tilde\psi'$.

The map $T$ is an {\it area-preserving diffeomorphism}. It is important
that $T$ is defined {\it not} in the whole $\mathbb{R}^2$,  but only on
the set of $\pi$-non-collapsing forward points for Eq.(\ref{1D_rep}),
i.e. ${\cal U}_\pi^+$. Inverse map $T^{-1}$ is defined on the set
${\cal U}_\pi^-$. Evidently, for each $p\in {\cal U}_\pi^+$ the image
$Tp\in{\cal U}_\pi^-$ and for each $q\in {\cal U}_\pi^-$ the pre-image
$T^{-1}q\in{\cal U}_\pi^+$, therefore $T{\cal U}_\pi^+={\cal U}_\pi^-$
and $T^{-1}{\cal U}_\pi^-={\cal U}_\pi^+$.

If, in addition, the potential $U(x)$ is {\it even}, $U(x)=U(-x)$,
Eq.(\ref{1D_rep}) is reversible. The prototypical example is the cosine
potential (\ref{cos-pot}) which appears as a basic model in numerous
studies. Denote $I$ the reflection with respect to $\psi$ axis in the
plane $(\psi,\psi')$. Due to reversibility of Eq.(\ref{1D_rep}), if
$p\in {\cal U}_\pi^+$ then
\begin{eqnarray}
T^{-1}Ip=ITp.\label{revers}
\end{eqnarray}
Therefore the sets ${\cal U}_\pi^+$ and ${\cal U}_\pi^-$ are connected
by the relations $I{\cal U}_\pi^+={\cal U}_\pi^-$, $I{\cal
U}_\pi^-={\cal U}_\pi^+$. The set ${\cal U}_\pi={\cal U}_\pi^+\cap{\cal
U}_\pi^-$ consists of the points which have both $T$-image and
$T$-pre-image. Theorem \ref{singular}.1 implies that ${\cal U}_\pi$ is
bounded. It follows from Sect.\ref{Ex:cos_pot} that ${\cal U}_\pi$ may
consist of several disjoined components $D_i$, $i=1,\ldots,N$.

The orbits defined by $T$ are sequences of points (finite, infinite or
bi-infinite) $\{p_n\}$, such that $Tp_{n}=p_{n+1}$. The fixed points of
$T$ correspond to $\pi$-periodic solutions of Eq.(\ref{1D_rep}) (such
solutions do exist for quite general periodic potential $U(x)$, see
\cite{Torres06}). For a fixed point $p$ let us denote $DT_p$ the
operator of linearization of $T$ at $p$. Let $\lambda_{1,2}$ be the
eigenvalues of $DT_p$. Since the map $T$ is area-preserving,
$\lambda_1\lambda_2=1$. Depending on the behavior of $T$ in a vicinity
of a fixed point, it may be of {\it elliptic} or {\it hyperbolic} type
\cite{Wiggins}. In the case of hyperbolic fixed point both
$\lambda_{1,2}$ are real and in the case of elliptic point they are
complex conjugated, $|\lambda_{1,2}|=1$. Also we call a {\it $k$-cycle}
an orbit which consists of points $p_1,\ldots,p_k\in \mathbb{R}^2$ such
that
\begin{eqnarray*}
Tp_1=p_2,\quad Tp_2=p_3,\ldots , \quad  Tp_k=p_1.
\end{eqnarray*}
Evidently $p_1,\ldots,p_k$ are fixed points for $T^k$. The $k$-cycles
correspond to $k\pi$-periodic solutions of Eq.(\ref{1D_rep}). A
$k$-cycle also may be of {\it elliptic} or {\it hyperbolic} type. This
is determined by the type (elliptic or hyperbolic) of the fixed point
$p_1$ for the map $T^k$.

Below we consider bi-infinite orbits which lie completely within the
set ${\cal U}_\pi$. Basing on Theorem \ref{SymDynTheory}.1 we formulate
necessary conditions which guarantee that these orbits can be coded
unambiguously by the sequences of numbers $i$ of $D_i$ in the order the
orbit ``visits'' them.

\subsection{Symbolic dynamics: application to Eq.(\ref{1D_rep})}\label{SymDynPract}

The application of Theorem \ref{SymDynTheory}.1 to Eq.(\ref{1D_rep})
gives  sufficient conditions for existence of  coding homeomorphism.
They can be formulated as follows:
\medskip

{\bf Hypothesis 1.} The set ${\cal U}_\pi$ consists of $N$ disjoined
islands $D_i$, $i=1,\ldots,N$, i.e. of $N$ curvilinear quadrangles
bounded by curves which possess some monotonic properties (see the
definitions in Sect.\ref{SymDynTheory}).
\medskip

{\bf Hypothesis 2.} The Poincare map $T$ associated with
Eq.(\ref{1D_rep}) is such that
\begin{itemize}
\item[(a)] $T$ maps v-strips of any $D_i$,
$i=1,\ldots,N$, in such a way that for any v-strip $V$, $V\in D_i$, all
the intersections $TV\cap D_j$, $j=1,\ldots,N$, are nonempty and
are v-strips.%
\item[(b)]  the inverse
map $T^{-1}$ maps h-strips of any $D_i$, $i=1,\ldots,N$, in such a way
that for any h-strip $H$, $H\in D_i$, the intersections $T^{-1}H\cap
D_j$, $j=1,\ldots,N$, are nonempty and are h-strips.
\end{itemize}
\medskip

{\bf Hypothesis 3.} The sequences of sets $\Delta_n^\pm$ defined as
follows
\begin{eqnarray*}
&&\Delta_0^+={\cal U}_\pi,\quad \Delta_n^+=T\Delta_{n-1}^+\cap {\cal
U}_\pi,\\
&&\Delta_0^-={\cal U}_\pi,\quad \Delta_n^-=T^{-1}\Delta_{n-1}^-\cap
{\cal U}_\pi,
\end{eqnarray*}
are such that $\lim_{n\to\infty}\mu(\Delta_n^\pm)=0$.\medskip

It follows from Theorem  \ref{SymDynTheory}.1 that if Hypotheses 1-3
hold then one can assert that there exists a homeomorphism  between all
bounded in $\mathbb{R}$ solutions of Eq.(\ref{1D_rep}) and the
sequences from $\Omega^N$ which can be regarded as {\it codes} for
these solutions. The verification of Hypotheses 1-3 can be done
numerically. From practical viewpoint, the following comments may be
useful:\medskip

1. If the periodic potential $U(x)$ is even the point (b) of Hypothesis
2 follows from the point (a). In fact, if $H$ is an h-strip then $IH$
is a v-strip where $I$ is a reflection with respect to $\psi$ axis.
Then the statement (b) follows from the relation
(\ref{revers}).\medskip

2. Let $DT_p$ be the operator of linearization of $T$ at point $p$. Let
\begin{eqnarray}
{\bf e}_1=\left(
\begin{array}{c}
1\\0
\end{array}
\right);\quad {\bf e}_2=\left(
\begin{array}{c}
0\\1
\end{array}
\right)\label{UnitV}
\end{eqnarray}
Define the functions
\begin{eqnarray*}
&&g_1(p)=(DT_{p}{\bf e}_1,{\bf e}_1)\cdot(DT_{p}{\bf e}_2,{\bf
e}_1),\\
&& g_2(p)=(DT_{p}{\bf e}_1,{\bf e}_2)\cdot(DT_{p}{\bf e}_2,{\bf e}_2).
\end{eqnarray*}
Then the following statement is valid:\medskip

{\bf Theorem \ref{CodingGen}.1.} {\it Assume that the potential $U(x)$
is even and the following conditions hold:
\begin{itemize}
\item ${\cal U}^+_\pi$ is an infinite curvilinear
strip;
\item ${\cal U}^+_\pi\cap{\cal U}^-_\pi={\cal U}_\pi=\bigcup_{i=1}^N D_i$ where $D_i$ are
non-overlapping islands;
\item for each pair $(i,j)$, if
\begin{itemize}
\item $\beta^\pm_i$ are graphs of monotone non-decreasing functions then for
any $p\in T^{-1}D_j\cap D_i$ the relations $g_1(p)>0$, $g_2(p)>0$ hold;
\item $\beta^\pm_i$ are graphs of monotone non-increasing functions then for
any $p\in T^{-1}D_j\cap D_i$ the relations $g_1(p)<0$, $g_2(p)<0$ hold.
\end{itemize}
\end{itemize}
Then the conditions of Hypothesis 2 take place.}\medskip

The proof of Theorem \ref{CodingGen}.1 can be found in \ref{ProofHyp2}.
It follows from Theorem \ref{CodingGen}.1 that numerical evidence for
Hypothesis 2 can be given by calculation of $g_1(p)$ and $g_2(p)$
within the set ${\cal U}_\pi$.\medskip

3. If the periodic potential $U(x)$ is even the relation (\ref{revers})
implies that $\mu(\Delta_n^+)=\mu(\Delta_n^-)$ for any $n$. Therefore
in order to verify Hypothesis 3 in this case it is enough to check the
condition $\lim_{n\to\infty}\mu(\Delta_n^+)=0$ only.\medskip

 In the next section we describe the results of
numerical study for Eq.(\ref{1D_rep}) with cosine potential
(\ref{cos-pot}). We present numerical evidence that Hypotheses 1-3 hold
for vast areas in the plane of parameters $(\omega,A)$.

\section{The case of cosine potential}\label{Cosine}

For numerical study of Eq.(\ref{1D_rep}) with cosine potential
(\ref{cos-pot}), i.e. of Eq.(\ref{1Dcos}),  special interactive
software was elaborated. It is aimed to fulfill thorough numerical
scanning of the plane $(\psi,\psi')$ of initial data and visualize the
sets ${\cal U}_L^\pm$ and ${\cal U}_L$ for a given $L$. Also the
software allows to measure areas of ${\cal U}_L$, to trace orbits
generated by iterations of $T$, to find fixed points of $T^k$,
$k=1,2,\ldots$, to calculate values $g_{1,2}(p)$ and visualize areas
where $g_{1,2}(p)>0$ and $g_{1,2}(p)<0$, and has some other useful
features.

For the numerical scanning of the plane  $(\psi,\psi')$ a 2D grid with
steps $\Delta\psi$, $\Delta\psi'$ was introduced.  For each initial
data $\psi(0)=\psi_0$, $\psi'(0)=\psi'_0$ at the grid nodes the Cauchy
problem for Eq.(\ref{1Dcos}) on the interval $[0;L]$ was solved
numerically. If the solution $\psi(x)$ of the Cauchy problem remains
bounded (in modulus) by some large number $B$ on the interval $[0;L]$
we concluded that no collapse occurs and these initial data were
regarded as an $L$-non-collapsing point. Typically the values
$\Delta\psi=0.0005$ and $\Delta\psi'=0.0002$ were taken. We found that
the results for $B=100$ and $B=1000$ in all the cases were almost
indistinguishable.

Let us set forth the results of the numerical study of Eq.(\ref{1Dcos})
for each of the Hypotheses separately.


\subsection{Hypothesis 1}\label{Hyp1}

Some examples of the sets ${\cal U}_\pi^+$ and ${\cal U}_\pi^-$ are
shown in Fig.\ref{U_pi}. In all the cases the sets ${\cal U}_\pi^+$ and
${\cal U}_\pi^-$ are {\it curvilinear strips}. We found that this is
{\it a general feature} of Eq.(\ref{1Dcos}) for all values of the
parameters $\omega$ and $A$ that we considered.

The shape of the strips ${\cal U}_\pi^\pm$ may be quite complex and
their intersection ${\cal U}_\pi$ may consist of different number of
disjoined sets. Since the strips ${\cal U}_\pi^+$ and ${\cal U}_\pi^-$
are related to each other by symmetry with respect to the $\psi$ axis,
the typical situation is that ${\cal U}_\pi$ consists of {\it several
number of curvilinear deltoids} (see Fig.\ref{U_pi}, panels B and D)
which are symmetrical with respect to $\psi$ or $\psi'$ axes.

\begin{figure}
\centerline{\includegraphics [scale=1]{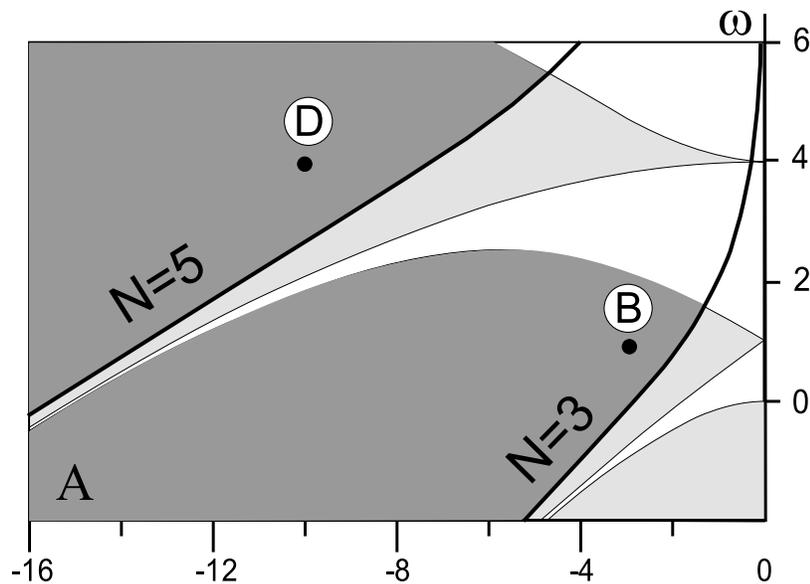}} \caption{The
plane of parameters $(\omega,A)$ with band and gap zones. The
boundaries of the regions where ${\cal U}_\pi$ is disjoined in three
(the curve $N=3$) or five (the curve $N=5$) connected components are
shown. These components are islands (in the sense of
Sect.\ref{SymDynTheory}) if $\omega$ and $A$ belong to the dark zones
between the marked curves and upper boundaries of the corresponding
gaps.} \label{IslZones(gen)}
\end{figure}

Fig.\ref{IslZones(gen)} shows the regions in the parameter plane
$(\omega,A)$ where such decomposition of ${\cal U}_\pi$ takes place.
Due to the symmetry
\begin{eqnarray*}
A\to-A,\quad x\to x+\pi/2
\end{eqnarray*}
the study has been restricted to the area $A<0$. The zones
corresponding to {\it gaps} and {\it bands} are also shown. Let us
remind that the separation of the gap zones and the band zones in the
parameter plane $(\omega,A)$ is  the key point for the theory of
linearized (Mathieu) equation \cite{AbrSteg}
\begin{equation}\label{Mathieu}
\psi_{xx}+(\omega-A\cos 2x)\psi=0
\end{equation}
If a point $(\omega,A)$ belongs to a band, all the solutions of
Eq.(\ref{Mathieu}) are bounded in $\mathbb{R}$ and if it belongs to a
gap, all of them are unbounded. It is known that  band/gap structure
also plays an important role in the theory of nonlinear equation
(\ref{1Dcos}) (see e.g. \cite{KonotSurvey}). In terms of the Poincare
map $T$ associated with Eq.(\ref{1Dcos}), if the point $(\omega,A)$ is
situated in a band, then the origin $O(0,0)$ is an elliptic fixed point
for $T$, and if $(\omega,A)$ lies in a gap then $O(0,0)$ is a
hyperbolic fixed point for $T$.

In Fig.\ref{IslZones(gen)} two curves marked  as $N=3$ and $N=5$ are
depicted. In the area above the curve $N=3$ and below the curve $N=5$
the set ${\cal U}_\pi$ consists of three connected components, in the
area above the curve $N=5$ it consists of five connected components
etc. Due to Theorem \ref{singular}.2 the boundary of each of the
component is continuous, but a conclusion about monotonicity and
Lipschitz properties of the boundaries should be made using numerical
arguments. Our numerical study indicates that {\it all these components
are islands} in the sense of Sect.\ref{SymDynTheory} in the areas (in
dark) between the marked curves and upper boundaries of the gaps, see
Fig.\ref{IslZones(gen)}.

We note that possible numbers of islands are related (indirectly) to
numbers of fixed points of the Poincare map $T$. In its turn, the
number of fixed points of $T$ is determined by the number of band or
gap where the point $(\omega,A)$ is situated. More detailed analysis of
these relations is an interesting issue for a further study.

\subsection{Hypothesis 2}\label{Hyp2}

Since the potential $U(x)$ is even and ${\cal U}^+_\pi$ and ${\cal
U}^-_\pi$ are infinite curvilinear strips, the verification of
Hypothesis 2 can be fulfilled using Theorem \ref{CodingGen}.1. To this
end the calculation of the values of $g_1(p)$ and $g_2(p)$ was
incorporated into the procedure of the numerical scanning 1 described
above. To confirm the results we also used direct visualization of the
vectors $DT_{p}{\bf e}_{1,2}$ for various $p\in{\cal U}_\pi$.

\begin{figure}
\centerline{\includegraphics [scale=0.8]{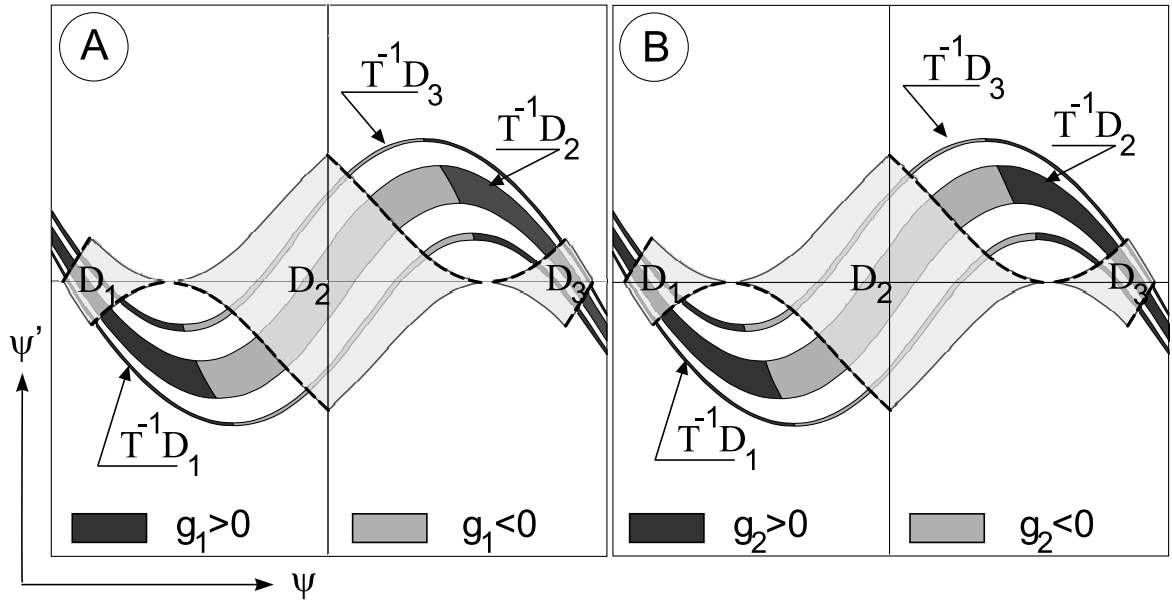}} \caption{The
region of the plane $(\psi,\psi')$, $-1.5<\psi<1.5$, $-1.5<\psi<1.5$
with the sets ${\cal U}_\pi$ and $T^{-1}{\cal U}_\pi$, $\omega=1$,
$A=-2$. The set ${\cal U}_\pi$ consists of three islands $D_1$, $D_2$
and $D_3$, the boundaries $\beta^\pm_{1,2,3}$ of the islands are marked
by bold dash lines. The areas where $g_1(p)>0$ and $g_1(p)<0$ for $p\in
T^{-1}D_{1,2,3}$ are indicated in panel A. Corresponding areas for
$g_2(p)$ are shown in panel B.} \label{g1g2}
\end{figure}

Let us describe in detail the case $\omega=1$, $A=-2$ which is typical
of the grey zone situated in the first gap, see
Fig.\ref{IslZones(gen)}. The set ${\cal U}_\pi$ consists of three
islands $D_1$, $D_2$ and $D_3$, see Fig.\ref{g1g2}. Their pre-images
$T^{-1}D_1$, $T^{-1}D_2$ and $T^{-1}D_3$ intersect $D_1$, $D_2$ and
$D_3$. Fig.\ref{g1g2} shows the signs of $g_{1,2}(p)$ for all the
intersections $T^{-1}D_i\cap D_j$, $i,j=1,2,3$. The boundaries
$\beta^\pm_{1,2,3}$ of the islands  are marked by bold dash lines. In
the islands $D_1$ and $D_3$ the boundaries $\beta^\pm_{1,3}$ are graphs
of increasing functions whereas for $D_2$ the boundaries
$\beta^\pm_{2}$ are  graphs of decreasing functions. It follows from
Fig.\ref{g1g2} that the signs of $g_{1,2}(p)$ conform the Hypothesis 2.

Overall, the numerical study shows that Hypothesis 2, as well as
Hypothesis 1, holds for $\omega$ and $A$ lying in the dark areas in
Fig.\ref{IslZones(gen)}.

\subsection{Hypothesis 3}\label{Hyp3}

\begin{figure}
\centerline{\includegraphics [scale=0.7]{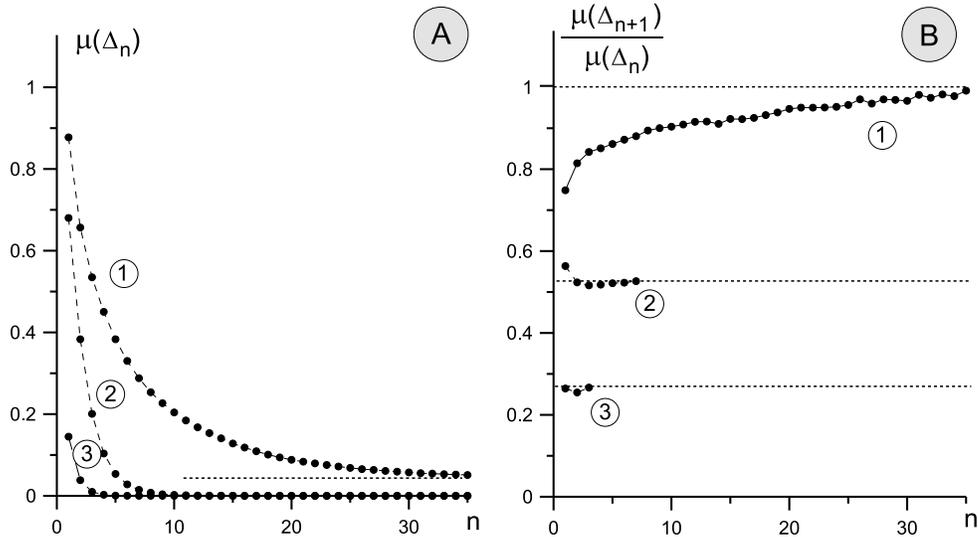}} \caption{Panel
(A): The area of the set $\Delta_n^+$ versus $n$ for $\omega=1$ and (1)
$A=-0.7$, (2) $A=-1.2$ and (3) $A=-2$. Panel (B): The ratio
$R_n=\mu(\Delta_{n+1}^+)/\mu(\Delta_{n}^+)$ for the same values of
$\omega$ and $A$.  Only few points are shown in panel (B) for the cases
(2) and (3)  since accuracy of calculations drastically falls due to
division of small numbers.} \label{Area}
\end{figure}

The behavior of $\mu(\Delta_n^+)$, $n=1,2,\ldots$, for various values
of parameters $\omega$ and $A$ was also studied numerically. Some of
the results are depicted in Fig.\ref{Area}.

It follows from Fig.\ref{Area} that Hypothesis 3 is valid not for all
the cases under consideration. A natural obstruction for Hypothesis 3
to hold is  presence of elliptic fixed points or cycles. Due to KAM
theory, in vicinity of an elliptic fixed point (or cycle) there exists
a set of positive measure that consists of points which remain in this
vicinity after any number of iterations of $T$. This means that
Hypothesis 3 is not valid if the point $(\omega,A)$ is situated in a
band in the plane of parameters (see Sect.\ref{Hyp1}), because in this
case the point $O(0,0)$ is an elliptic fixed point of $T$. This
situation takes place for the case 1, $\omega=1$ and $A=-0.7$, in
Fig.\ref{Area}.

As the point $(\omega,A)$ crosses a lower boundary of a gap in the
plane of parameters, the point $O(0,0)$ becomes a hyperbolic fixed
point and a pair of elliptic 2-cycles appears. It have been observed
that this bifurcation is the first bifurcation in a cascade of period
doubling bifurcations. Each of the bifurcations of this cascade gives
birth to elliptic cycles of double period. Omitting the details, we
summarize that the gap zones in the plane $(\omega,A)$ also contain
areas where Hypothesis 3 is not valid due to presence of the elliptic
cycles.

At the same time, if the point $(\omega,A)$ is situated in the dark
zones in Fig.\ref{IslZones(gen)}, numerical results show that
Hypothesis 3 holds. Moreover, our results allow to suppose {\it
exponential convergence} of $\mu(\Delta_n^+)$ to zero. The ratios
$R_n=\mu(\Delta_{n+1}^+)/\mu(\Delta_{n}^+)$ are shown in panel B of
Fig.\ref{Area}. For the cases 2 and 3, these ratios are smaller then 1
and remain close to the value $\mu(T{\cal U}_\pi\cap {\cal
U}_\pi)/\mu({\cal U}_\pi)$.

To summarize, basing on the numerical results presented above one can
conclude that {\it if $\omega$ and $A$ are situated in the dark zones
of the parameter plane, see Fig.\ref{IslZones(gen)}, the conditions of
Hypotheses 1-3 hold}. Therefore for these values of parameters all the
nonlinear states of GPE with cosine potential can be put in one-to-one
correspondence with codes from $\Omega^N$. When crossing the lower
boundary of the grey zones (marked $N=3$ or $N=5$ in
Fig.\ref{IslZones(gen)}) the conditions of Hypotheses 1 fail whereas
other two Hypotheses remain valid.

\subsection{More details about the map $T$}\label{Detail}

Let us describe in more detail the transformation of the sets ${\cal
U}_\pi^+$ and ${\cal U}_\pi$ by the map $T$.

\begin{figure}
\centerline{\includegraphics [scale=0.7]{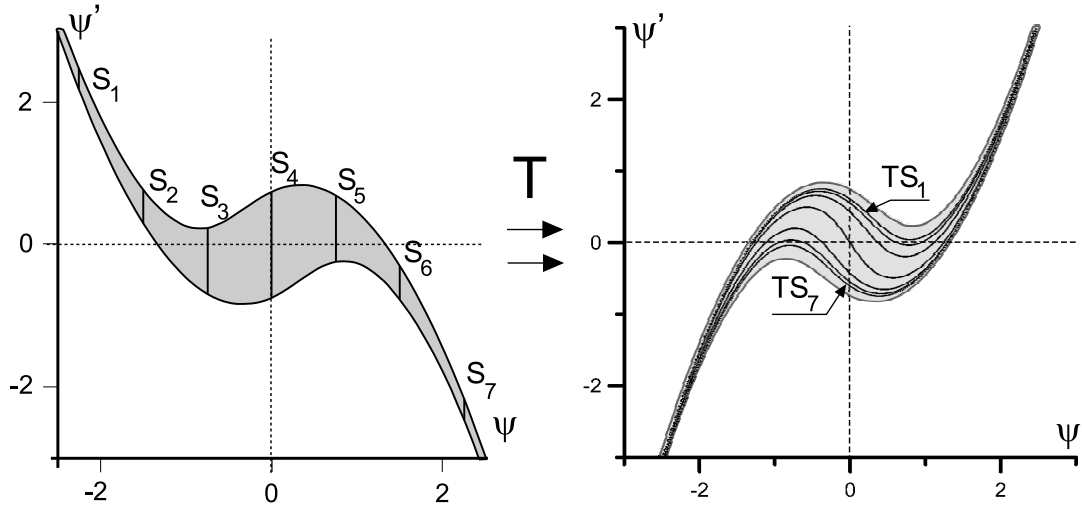}}\bigskip
\caption{The action of $T$ on ${\cal U}_\pi^+$, $\omega=1$, $A=-1.5$.
Vertical sections $S_k$, $k=1\div 7$ of ${\cal U}_\pi^+$ are mapped
into the infinite curves shown in the right panel.} \label{Trans}
\end{figure}

\begin{figure}
\centerline{\includegraphics [scale=0.7]{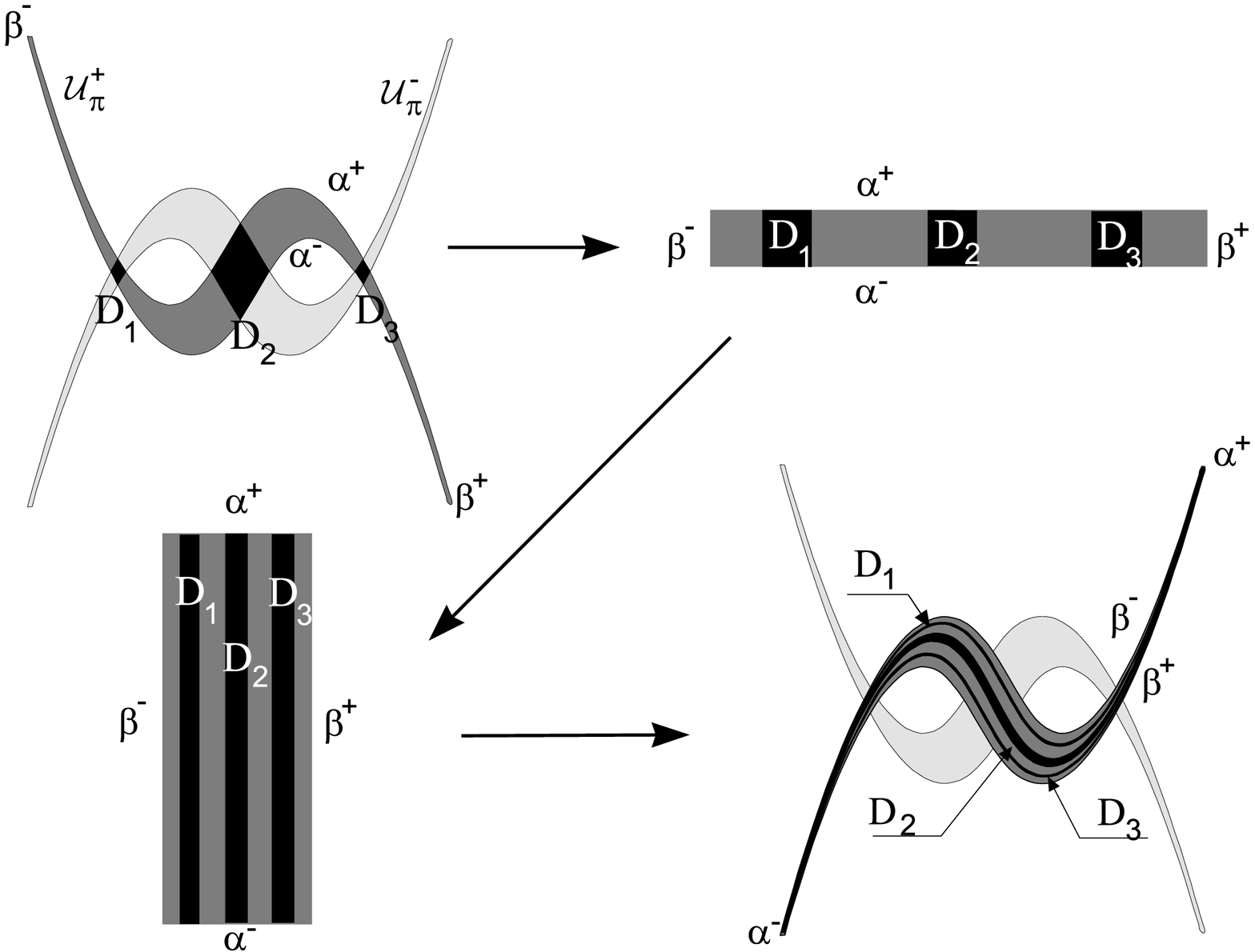}} \caption{The
action of the map $T$ on ${\cal U}_\pi^+$ and the islands $D_1$, $D_2$,
$D_3$ (the islands and their images are indicated by the same letters).
The map $T$ can be regarded as a composition of three transformations:
${\cal U}_\pi^+\to S_1\to S_2\to {\cal U}_\pi^-$: (i) deformation of
${\cal U}_\pi^+$ into infinite strip $S_1$, the ``boundaries''
$\beta^\pm$ of both, ${\cal U}_\pi^+$ and $S_1$, are in infinity; (ii)
stretching of $S_1$ in one dimension and contraction in another one to
get an infinite strip $S_2$, the boundaries $\beta^\pm$ become vertical
lines but $\alpha^\pm$ go to infinity; (iii) deformation of $S_2$ into
${\cal U}_\pi^-$. As a result, $T$ maps the islands $D_1$, $D_2$, $D_3$
into infinite curvilinear strips. Each of these strips crosses again
the islands $D_1$, $D_2$, $D_3$ and each of the intersections is a
v-strip. The shape of ${\cal U}_\pi^+$ was calculated for $\omega=1$,
$A=-3$.} \label{StrataMap}
\end{figure}

The action of $T$ on vertical sections of ${\cal U}_\pi^+$ is shown in
Fig.\ref{Trans}. Note that $T{\cal U}_\pi^+={\cal U}_\pi^-=I{\cal
U}_\pi^+$ where $I$ is the reflection with respect to $\psi$ axis.
Fig.\ref{StrataMap} illustrates the mapping of ${\cal U}_\pi^+$ and the
islands $D_1$, $D_2$, $D_3$ (the shape of ${\cal U}_\pi^+$ was
calculated for $\omega=1$, $A=-3$). It is practical to represent the
map $T$ as a composition of three transformations: ${\cal U}_\pi^+\to
S_1\to S_2\to {\cal U}_\pi^-$, where $S_1$ is an infinite horizontal
strip and $S_2$ is an infinite vertical strip, see Fig.\ref{StrataMap}.
The ``boundaries'' $\beta^\pm$ of both, ${\cal U}_\pi^+$ and $S_1$, are
in infinity. The transformation ${\cal U}_\pi^+\to S_1$ is a
deformation. The transformation of $S_1$ to $S_2$ consists in
stretching of $S_1$ in one dimension and contraction in another in such
a way that the boundaries $\beta^\pm$ transform into vertical lines but
$\alpha^\pm$ go to infinity. The transformation $S_2\to {\cal U}_\pi^-$
is again a deformation. As a result, $T$ maps the islands $D_1$, $D_2$,
$D_3$ into infinite curvilinear strips. Each of these strips crosses
the islands $D_1$, $D_2$, $D_3$ and each of the intersections is a
v-strip.

\begin{figure}
\centerline{\includegraphics [scale=0.7]{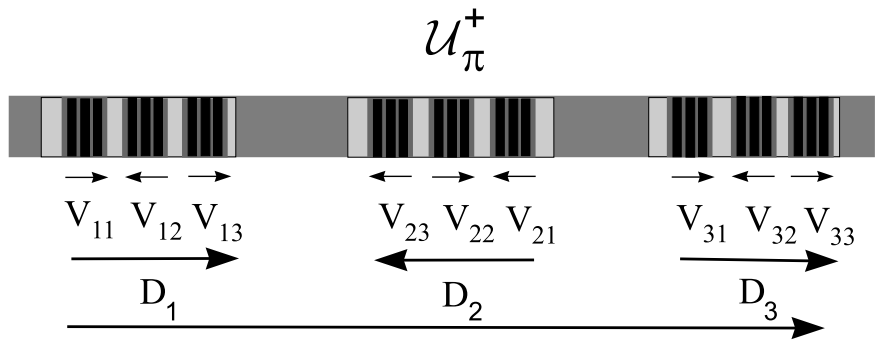}}
\caption{Algorithm for ordering of v-strips in ${\cal U}_\pi^+$.}
\label{Arrows}
\end{figure}

An interesting issue  is the study of {\it ordering} of the v-strips in
${\cal U}_\pi^+$ and the h-strips in ${\cal U}_\pi^-$ corresponding to
codes with coinciding blocks. The understanding of the strip ordering
is also of practical usage, since it explains the order of the
nonlinear modes as they appear in shooting procedure, see
Sect.\ref{Examples}. We describe the ordering of v-strips, the ordering
of h-strips is similar. Assume that ${\cal U}_\pi$ consists of $N$
disjoined islands and $N$ is odd. Consider orbits that visit the
islands $D_{i_{-n}},D_{i_{-n+1}},\ldots,D_{i_0}$, in the given order.
The points in $D_{i_0}$ which has this ``prehistory'' are situated in a
strip $V_{i_{-n}i_{-n+1}\ldots i_{-2}i_{-1}i_0}$ constructed by the
following recurrence rule
\begin{eqnarray*}
&&V_{i_{-n}i_{-n+1}}=TD_{i_{-n}}\cap D_{i_{-n+1}}\\
&&\ldots\\
&&V_{i_{-n}i_{-n+1}\ldots i_{-2}i_{-1}i_0}=TV_{i_{-n}i_{-n+1}\ldots
i_{-2}i_{-1}}\cap D_{i_0}.
\end{eqnarray*}
and (see \ref{ProofMain})
\begin{eqnarray*}
\ldots\subset V_{i_{-n}i_{-n+1}\ldots i_{-2}i_{-1}i_0}\subset
V_{i_{-n+1}\ldots i_{-2}i_{-1}i_0}\subset\ldots \subset
V_{i_{-2}i_{-1}i_0}\subset V_{i_{-1}i_0}\subset D_{i_0}
\end{eqnarray*}
The orbit of a point $p\in V_{i_{-n}i_{-n+1}\ldots i_{-2}i_{-1}i_0}$
has in its code a block
\begin{eqnarray*}
(\ldots \underbrace{i_{-n}i_{-n+1}\ldots i_{-2}i_{-1}i_0}\ldots).
\end{eqnarray*}
Since each of $i_k$ can take the values $1,\ldots,N$, there are
$N^{n+1}$ strips in ${\cal U}_\pi^+$ each coded by the sequence of
length $n+1$. The algorithm for their ordering in ${\cal U}_\pi^+$
follows immediately from geometrical properties of the intersection of
the strips ${\cal U}_\pi^-$ and ${\cal U}_\pi^+$. It can be described
as follows:
\medskip

1. Mark the islands $D_1$, $\ldots$, $D_N$  as they are ordered in
${\cal U}_\pi^+$, see Fig.\ref{Arrows}. Draw an arrow $I_0$ over all of
them pointing from $D_1$ to $D_N$.

2. Draw arrows $I_{01},\ldots,I_{0N}$ over each island in such a way
that the directions of rightmost $I_{01}$ and leftmost $I_{0N}$ arrows
coincided with the direction of $I_0$, but the directions of any two
neighboring arrows were opposite. Sketch v-strips in $D_{i_0}$ in such
a way that their ordering (from $V_{1 i_0}$ to $V_{N i_0}$) agreed with
the direction of the arrow $I_{0 i_0}$.

3. Draw an arrows over each of $V_{i_{-1} i_0}$ by the same manner and
sketch v-strips $V_{i_{-2}i_{-1} i_0}$ according the directions of
these arrows, etc.

It turns out that the algorithm given above is similar to procedure of
ordering of localized modes for DNLS described in \cite{AlfBraKon}.

\subsection{Examples}\label{Examples}

Let the parameters $(\omega,A)$ belong to the dark zone in the first
gap, see Fig.\ref{IslZones(gen)}. Then all the bounded in $\mathbb{R}$
solutions of Eq.(\ref{1Dcos}) can be coded by bi-infinite sequences of
three symbol ``alphabet''. Conversely, for each bi-infinite ``word''
composed by symbols of this ``alphabet'' these exists a solution with
corresponding code. The symbols may be chosen as ``$-$'', ``$0$'' and
``$+$'', and they mark entering of the orbit into $D_1$, $D_2$ and
$D_3$ respectively.\medskip

\begin{figure}
\centerline{\includegraphics [scale=0.9]{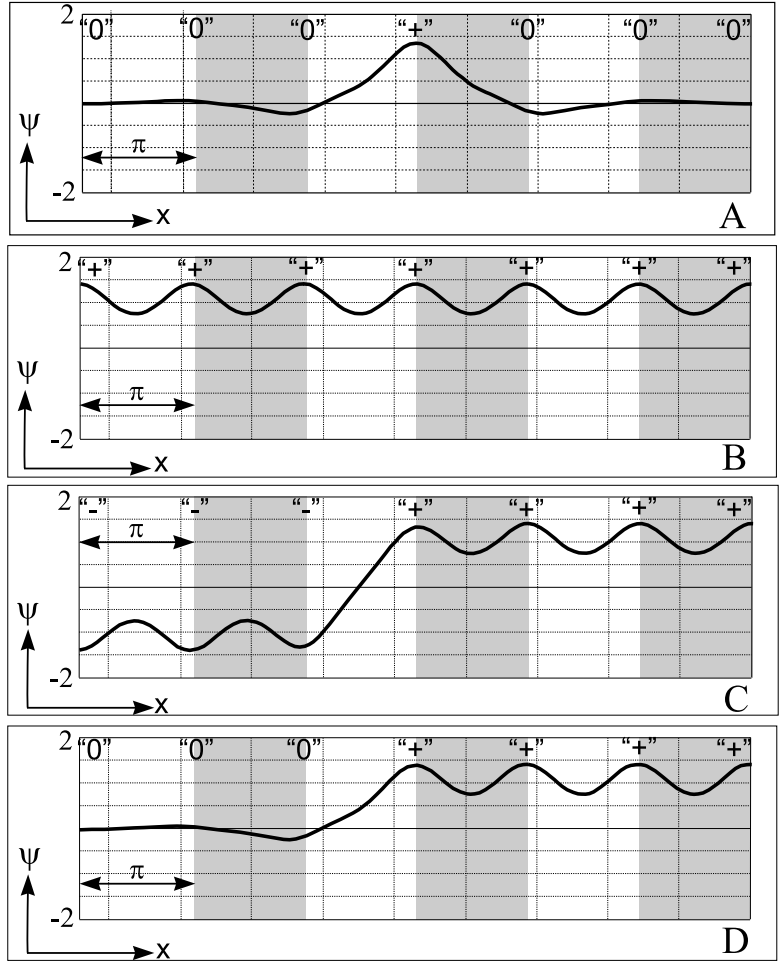}}
\caption{Nonlinear states for GPE, $\omega=1$, $A=-2$: A: bright
gap-soliton, the code $(\ldots00+00\ldots)$; B: $\pi$-periodic
structure, the code $(\ldots+++\ldots)$; C: dark soliton, the code
$(\ldots---+++\ldots)$; D: domain wall, the code
$(\ldots000+++\ldots)$.} \label{BasShapes}
\end{figure}

{\it Example 1.} Localized modes described by Eq.(\ref{1Dcos})
correspond to the codes with finite numbers of nonzero symbols. In
particular, Eq.(\ref{1Dcos}) admits well-known solution in the form of
bright gap-soliton, $\psi(x)$, \cite{Kiv2003,AKS}, localized in one
well of the potential, see Fig.\ref{BasShapes} A. This solution
corresponds to the code $(\ldots00+00\ldots)$. Also there exist the gap
soliton solution $-\psi(x)$ with the code $(\ldots00-00\ldots)$.
\medskip

{\it Example 2.} There exist exactly two $\pi$-periodic solutions of
Eq.(\ref{1Dcos})  with the codes $(\ldots+++\ldots)$ and
$(\ldots---\ldots)$, related to each other by symmetry $\psi\to-\psi$,
see Fig.\ref{BasShapes} B.\medskip

{\it Example 3.} There exists dark soliton solution of Eq.(\ref{1Dcos})
corresponding to the code $(\ldots---+++\ldots)$, see
Fig.\ref{BasShapes} C. Also the coding predicts that there exist other
solutions of this type, having the codes $(\ldots---0+++\ldots)$,
$(\ldots---00+++\ldots)$, etc.\medskip

{\it Example 4.} There exist ``domain wall''-type solutions
corresponding to the codes $(\ldots000+++\ldots)$,
$(\ldots---000\ldots)$. These objects were found to exist in the case
of GPE with attractive nonlinearity \cite{DomWalls05}. The coding
approach predicts their existence in the case of repulsive nonlinearity
also. They have been found numerically, see Fig. \ref{BasShapes} D.
\medskip

\begin{figure}
\centerline{\includegraphics [scale=1.05]{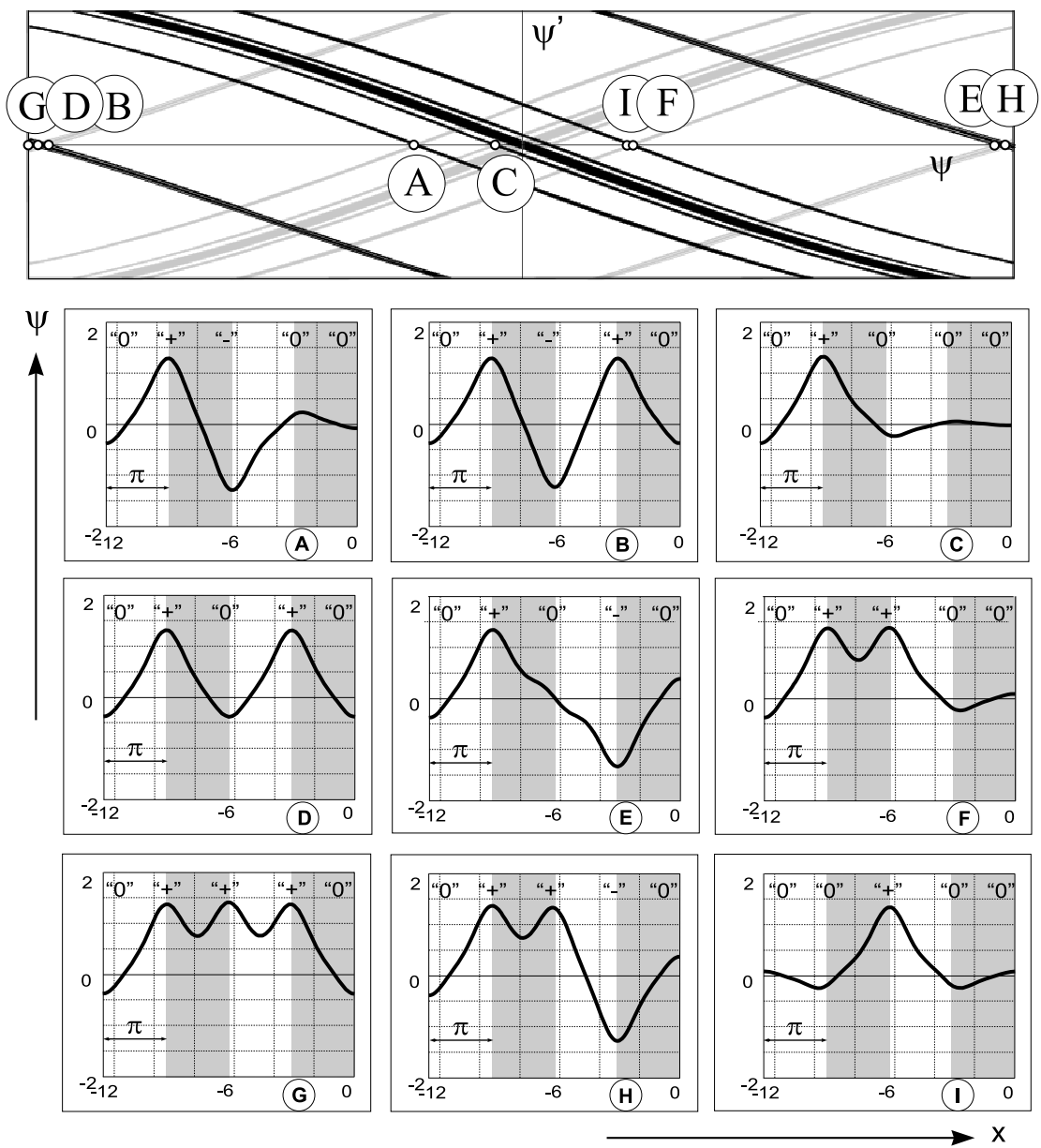}} \caption{The
solutions of Eq.(\ref{1Dcos}) ($A=-2$, $\omega=1$) on the interval
$[-4\pi,0]$ with Neumann boundary conditions at $x=-4\pi$ and $x=0$
(nine panels below) and the position of corresponding points in the
intersection of ${\cal U}_{4\pi}^+$ and ${\cal U}_{4\pi}^-$ (upper
panel). The upper panel represents the scaled rectangle marked in panel
D of Fig.\ref{Fractals} with fragments of ${\cal U}_{4\pi}^-$ (in
black) and ${\cal U}_{4\pi}^+$ (in grey). The codes (the blocks
$(\theta_1\theta_2\theta_3\theta_4\theta_5)$) are: A: $(0+-00)$; B:
$(0+-+0)$; C: $(0+000)$; D: $(0+0+0)$; E: $(0+0-0)$; F: $(0++00)$; G:
$(0+++0)$; H: $(0++-0)$; I: $(00+00)$.} \label{Codes}
\end{figure}

{\it Example 5.} Consider boundary value problem for Eq.(\ref{1Dcos})
on the interval  $[-4\pi,0]$ with Neumann boundary conditions at
$x=-4\pi$ and $x=0$. These solutions can be viewed as reductions to the
interval of length $4\pi$ of periodic solutions with period $8\pi$
which satisfy additional symmetry conditions
\begin{eqnarray*}
u(x)=u(-x);\quad u(-4\pi+x)=u(-4\pi-x).
\end{eqnarray*}
The codes for these solutions are of the form
\begin{eqnarray*}
(\ldots \theta_5\theta_4\theta_3\theta_2
\underbrace{\theta_1\theta_2\theta_3\theta_4\theta_5\theta_4\theta_3\theta_2}_{\mbox{the
period}}\theta_1\theta_2\theta_3\theta_4\theta_5\ldots),
\end{eqnarray*}
where $\theta_i$, $i=1\div5$, is one of the symbols ``$+$'', ``$0$'' or
``$-$''. Therefore there are $3^5=243$ solutions of this type. In
Fig.\ref{Codes} nine of these solutions are depicted for $A=-2$,
$\omega=1$.  From numerical viewpoint, these nonlinear modes can be
found by shooting method taking initial data $\psi(-4\pi)=\tilde\psi$,
$\psi_x(-4\pi)=0$ and adjusting $\tilde\psi$ in such a way that
$\psi_x(0)=0$. As $\tilde\psi$ increases, one gets the nonlinear modes
one by one in the order described in Sect.\ref{Detail}.

All the solutions shown in Fig.\ref{Codes} have the codes with
$\theta_1=\theta_5=$``$0$''.  Therefore they can be viewed as
approximations for localized modes which have the domain of
localization of length $4\pi$. These localized modes correspond to the
codes
\begin{eqnarray*}
(\ldots00~\theta_2\theta_3\theta_4~00\ldots).
\end{eqnarray*}
There are $3^3=27$ sequences of this type but only 10 of them
(including one which consists of zeroes only and corresponds to the
zero solution) are different in the sense that they are not related to
each other by symmetry reductions. Fig.\ref{Codes} shows just these
nine nonzero solutions.

\section{Conclusion}\label{concl}

In this paper, we describe the method for coding of nonlinear states
covered by 1D Gross-Pitaevskii equation with periodic potential $U(x)$
and repulsive nonlinearity. We prove that under certain conditions
there exists one-to-one correspondence between the set of all bounded
in $\mathbb{R}$ solutions of Eq.(\ref{1D_rep}) and the set of
bi-infinite sequences of numbers $1,\ldots,N$. These sequences can be
regarded as codes for the solutions of Eq.(\ref{1D_rep}).  The number
$N$ is determined by the parameters of Eq.(\ref{1D_rep}). It is
important that (i) each coding sequence corresponds to one and only one
solution and (ii) each solutions has corresponding code. The conditions
for the coding to be possible are presented in a form of three
hypotheses. For a given $U(x)$, the hypotheses should be verified
numerically. We report on numerical check of these hypotheses for the
case of cosine potential, i.e., for Eq.(\ref{1Dcos}), and indicate the
regions in the plane of parameters $(\omega,A)$ where the coding is
possible.

Heuristically, the coding technique described above can be interpreted
as follows. The periodic potential can be regarded as an infinite chain
of equidistantly spaced potential wells. It is known that if $U(x)$ is
a ``deep enough'' single-well potential, Eq.(\ref{1D_rep}) admits  one
or more  localized solutions called ``fundamental gap soliton'', FGS,
in \cite{Zhang09_1,Zhang09_2}. Also Eq.(\ref{1D_rep}) admits zero
state. Assume that in total there exist $N$ states (including zero
state) described by the single well potential. Note that the number $N$
is odd: since the nonlinearity is odd if $\psi(x)$ is a FGS then
$-\psi(x)$ is also a solution Eq.(\ref{1D_rep}) and the zero solution
should also be taken into account. In these terms the coding means that
one assigns to each of possible single-well states a number from 1 to
$N$ and attributes to ``bound states'' of these entities situated in
the wells of periodic potential bi-infinite ``words'' consisting of
numbers from 1 to $N$.  This viewpoint exploits an analogy between
periodic problem and discrete problem replacing the solution on each
period by a lattice node. The corresponding reduction can be made
consistently and rigorously using Wannier functions technique
\cite{Wannier02} but the resulting system of discrete equations is
nonlocal and quite difficult for a comprehensive study.

The approach presented in this paper may be applied to
Eq.(\ref{1D_rep}) with different types of the periodic potential
$U(x)$. Also it may be extended in various directions. In particular,
preliminary studies show that it can be applied with minor
modifications to the equation
\begin{equation}\label{Eq3-5}
\psi_{xx}+(\omega-U(x))\psi+g_1\psi^3+g_2\psi^5=0.
\end{equation}
where $g_2<0$. Eq.(\ref{Eq3-5}) also arises in the theory of BEC
\cite{AS05,AKP07}.  The shapes of the sets ${\cal U}^\pm_\pi$ in this
case are similar to ones described above for Eq.(\ref{1D_rep}). Another
possible extension of this approach can be made for {\it complex}
nonlinear states of GPE, $\psi(x)=\rho(x)\exp\{i\phi(x)\}$. It is known
\cite{AKS} that the amplitude $\rho(x)$ obeys the equation
\begin{eqnarray}
\rho_{xx}+(\omega-U(x))\rho-\rho^3-\frac{M^2}{\rho^3}=0, \label{Eq313}
\end{eqnarray}
where $M$ is an arbitrary real constant. For a given amplitude
$\rho(x)$, the phase $\phi(x)$ can be found from the relation
$\rho^2(x)\phi_x(x)=M$. Other possible extensions of the approach may
be related to the cases when ${\cal U}_\pi$ consists of partially
overlapping islands or of more general sets which are not islands at
all.

Having in hand a complete description of nonlinear modes for
Eq.(\ref{1D_rep}) in terms of their codes, one can return to the
problem of stability of these modes. In our opinion, a relation between
the code and the stability of corresponding mode is an interesting
issue for further study.
A good example of such a study is paper \cite{PelKevFr05} where the
similar problem was considered for DNLS.

At last, let us note that the approach developed in this paper cannot
be applied (at least, directly) to the case of GPE with attractive
interactions, $\sigma<0$ in Eq.(\ref{1D}). One can prove that in this
case all the solutions of Eq.(\ref{1D}) are non-collapsing under quite
general assumptions for the potential $U(x)$.

\ack

Authors are grateful to Dr. D.Zezyulin and P.Kizin for careful reading
of the manuscript making a lot of valuable comments. The work of
authors was supported by Russian federal program ``Scientific and
educational personnel of the innovative Russia'', grant 14.B37.21.1273.

\appendix


\section{Proof of Theorem \ref{singular}.1}\label{Th1}

The following statement proved in \cite{AlfZez} will be used
below:\medskip

{\bf Comparison Lemma.} {\it Let the functions $y(t)$ and $x(t)$,
$t\in[a;b]$ be solutions of equations
\begin{eqnarray}
&&y_{tt}-g(t,y)=0\label{LemEq1}\\
&&x_{tt}-f(t,x)=0\label{LemEq2}
\end{eqnarray}
correspondingly. Let also the following conditions hold:
\medskip

(i) $f(t,\xi)$, $g(t,\xi)$ are defined on $[a;b]\times [A;B]$ and
locally Lipschitz continuous with respect to $\xi$, $\xi\in [A;B]$,
($A$,$B$, $b$ may be finite or infinite);
\medskip

(ii) $g(t,\xi)\geq f(t,\xi)$ for any $t\in [a;b]$, $\xi\in [A;B]$;
\medskip

(iii) $f(t,\xi)$ is monotone nondecreasing with respect to $\xi$,
$\xi\in [A;B]$.
\medskip

Let $A<x(a)\leq y(a)<B$ and $x_t(a)\leq y_t(a)$. Then $x_t(t)\leq
y_t(t)$ and $x(t)\leq y(t)$ while $A<x(t),y(t)<B$ or for the whole
interval $t\in[a;b]$}
\medskip

In what follows we assume that the potential $U(x)$ be continuous and
bounded on $\mathbb{R}$ and use the notations introduced in
Sect.\ref{Def}. To prove Theorem \ref{singular}.1 we need the following
lemmas:
\medskip

{\bf Lemma \ref{singular}.1.} {\it Let $\psi'_0\geq 0$. Then for
$\psi_0>\sqrt{\overline{\Omega}}$
\begin{eqnarray}\label{Est_h}
h^+(\psi_0,\psi'_0)\leq
h^+_0(\psi_0)\equiv\int_{\psi_0}^\infty\frac{\sqrt{2}~d\eta}{\sqrt{\eta^4-\psi_0^4-2\overline{\Omega}(\eta^2-\psi_0^2)}},
\end{eqnarray}
and $h^+_0(\psi_0)\sim \frac{{\rm K}(1/\sqrt{2})}{\psi_0}$ when
$\psi_0\to+\infty$. Here ${\rm K}(\cdot)$ is complete elliptic integral
of the first kind.}
\medskip

{\it Proof:} Consider the equation
\begin{eqnarray}
\phi_{xx}+\overline{\Omega}\phi-\phi^3=0 \label{psi3}
\end{eqnarray}
The solution $\phi(x)$ for Eq.(\ref{psi3}) with initial data
$\phi(0)=\psi_0$, $\phi_x(0)=\psi_0'$ can be written in implicit form
as follows
\begin{eqnarray}\label{ExactPsi3}
x=\int_{\psi_0}^\phi\frac{\sqrt{2}~d\eta}{\sqrt{\Delta-2\overline{\Omega}\eta^2+\eta^4}};\quad
\Delta=2(\psi_0')^2+2\overline{\Omega}\psi_0^2-\psi_0^4.
\end{eqnarray}
The solution $\phi(x)$ tends to $+\infty$ at the point
\begin{eqnarray*}
x_0=\int_{\psi_0}^\infty\frac{\sqrt{2}~d\eta}{\sqrt{\eta^4-\psi_0^4-2\overline{\Omega}(\eta^2-\psi_0^2)+2(\psi_0')^2}}
\end{eqnarray*}
and the following estimation holds
\begin{eqnarray*}
x_0\leq
h^+(\psi_0)\equiv\int_{\psi_0}^\infty\frac{\sqrt{2}~d\eta}{\sqrt{\eta^4-\psi_0^4-2\overline{\Omega}(\eta^2-\psi_0^2)}}\sim\frac{{\rm
K}(1/\sqrt{2})}{\psi_0}, \quad \psi_0\to+\infty
\end{eqnarray*}
Now let us consider the solution $\psi(x)$ of Eq.(\ref{1D_rep}) with
initial data  $\psi(0)=\psi_0$, $\psi_x(0)=\psi_0'$. Since for
$\xi>\sqrt{\overline{\Omega}}$ the function
$F(\xi)=\xi^3-\overline{\Omega}\xi$ is monotonic and
$\xi^3-(\omega-U(x))\xi\geq F(\xi)$ one can apply Comparison Lemma from
\cite{AlfZez} to Eq.(\ref{1D_rep}) and Eq.(\ref{psi3}). Therefore for
$x>0$ the inequality $\psi(x)\geq\phi(x)$ holds. This means that
$\psi(x)$ collapses at a point $h^+(\psi_0,\psi'_0)\leq x_0\leq
h^+_0(\psi_0)$. This proves Lemma \ref{singular}.1. $\blacksquare$
\medskip

{\bf Lemma \ref{singular}.2.} {\it For each $L$ there exists a value
$\tilde{\psi}_L$ such that the set ${\cal U}_L$ is situated in the
plane $\mathbb{R}^2=(\psi,\psi')$ in the strip
$-\tilde{\psi}_L<\psi<\tilde{\psi}_L$.}\medskip

{\it Proof:} Due to Lemma \ref{singular}.1 for each $L$ there exists
$\tilde{\psi}_L$ such that there are no points of ${\cal U}_L$ in the
sector $\psi>\tilde{\psi}_L$, $\psi'\geq 0$. Since Eq.(\ref{1D_rep}) is
invariant with respect to the symmetry $\psi\to-\psi$ the estimation
(\ref{Est_h}) holds also for $\psi'_0\leq 0$ and
$\psi_0<-\sqrt{\overline{\Omega}}$, therefore there are no points of
${\cal U}_L$ in the sector $\psi<-\tilde{\psi}_L$, $\psi'\leq 0$.
Making the transformation $x\to-x$ and repeating the reasoning of Lemma
\ref{singular}.1 we obtain the estimation
\begin{eqnarray*}
h^-(\psi_0,\psi'_0)\leq
h^-_0(\psi_0)\equiv\int_{-\infty}^{\psi_0}\frac{\sqrt{2}~d\eta}{\sqrt{\eta^4-\psi_0^4-2\overline{\Omega}(\eta^2-\psi_0^2)}}
\end{eqnarray*}
for the two cases: (i) $\psi_0>\sqrt{\overline{\Omega}}$,
$\psi'_0\leq0$ and (ii) $\psi_0<-\sqrt{\overline{\Omega}}$,
$\psi'_0\geq0$. Similarly, $h^-_0(\psi_0)\sim-\frac{{\rm
K}(1/\sqrt{2})}{\psi_0}$, $\psi_0\to-\infty$. Therefore there are no
points of ${\cal U}_L$ in the sectors $\psi>\tilde{\psi}_L$, $\psi'\leq
0$ and $\psi<-\tilde{\psi}_L$, $\psi'\geq 0$. This implies the
statement of Lemma \ref{singular}.2. $\blacksquare$
\medskip

{\it Proof of Theorem \ref{singular}.1:}  Due to Lemma \ref{singular}.2
there exists the value $\tilde{\psi}_L$ such that no points of ${\cal
U}_L$ are situated out of the strip
$-\tilde{\psi}_L<\psi<\tilde{\psi}_L$. Therefore it is enough to prove
that there are no points of ${\cal U}_L$ in two half-strips
\begin{eqnarray*}
S^+_R=\{(\psi,\psi')\in\mathbb{R}^2|~
-\tilde{\psi}_L<\psi<\tilde{\psi}_L,~
\psi'>R\}\\
S^-_R=\{(\psi,\psi')\in\mathbb{R}^2|~
-\tilde{\psi}_L<\psi<\tilde{\psi}_L,~ \psi'<-R\}
\end{eqnarray*}
for $R$ large enough. Let us prove this fact for $S^+_R$, the proof for
$S^-_R$ is analogous.  It follows from Lemma \ref{singular}.2 that
there exists the value $\tilde{\psi}_{L/2}>\tilde{\psi}_{L}$ such that
all the points $(\psi,\psi')$ for $\psi>\tilde{\psi}_{L/2}$ and
$\psi'\geq0$ are $L/2$-collapsing forward points. Introduce the value
\begin{eqnarray*}
M_L\equiv
\min_{\begin{array}{c}\xi\in[-\tilde{\psi}_{L/2};\tilde{\psi}_{L/2}]\\\eta\in[\underline{\Omega};\overline{\Omega}]
\end{array} }\left(\xi^3-\eta \xi\right)
\end{eqnarray*}
Evidently $M_L\leq0$. Consider the solution $\psi(x)$ of Cauchy problem
for Eq.(\ref{1D_rep}) with initial data $\psi(0)=\psi_0$, $\psi_0\in
[-\tilde{\psi}_L;\tilde{\psi}_L]$ and $\psi_x(0)=\psi'_0$, $\psi'_0>R$.
While $-\tilde{\psi}_{L/2}\leq\psi(x)\leq \tilde{\psi}_{L/2}$ one has
$\psi_{xx}(x)\geq M_L$. Then the following relations hold
\begin{eqnarray}
&&\psi_x(x)\geq \psi_x(0)+M_Lx\geq R+M_Lx,\label{Est_psi'}\\[2mm]
&&\psi(x)\geq \psi(0)+Rx+\frac {M_L}2 x^2\geq-\tilde{\psi}_L+Rx+\frac
{M_L}2 x^2.\label{Est_psi}
\end{eqnarray}
We claim that if the initial data for Eq.(\ref{1D_rep}) are situated in
the half-strip $S^+_R$ with
\begin{eqnarray*}
R>\tilde{\psi}'_L\equiv\max\left\{-\frac{M_LL}2,
\frac1{4L}\left(8\tilde{\psi}_{L/2}+8\tilde{\psi}_L-M_LL^2\right)\right\}
\end{eqnarray*}
then the segment of curve $\{(\psi(x),\psi_x(x)),~0\leq x\leq L/2\}$
crosses the line $\psi=\tilde{\psi}_{L/2}$ in some point where
$\psi'\geq0$. In fact, assuming that
$-\tilde{\psi}_{L/2}\leq\psi(x)\leq \tilde{\psi}_{L/2}$ for
$x\in[0;L/2]$ from (\ref{Est_psi'}) and (\ref{Est_psi}) one concludes
that
\begin{eqnarray*}
\psi_x(x)&\geq& R+M_L x>0,\quad
x\in[0;L/2];\\[2mm]
\psi(0)&\leq&\tilde{\psi}_L\leq \tilde{\psi}_{L/2}; \quad\psi(L/2)\geq
-\tilde{\psi}_L+\frac {RL}2+\frac {M_LL^2}8>\tilde{\psi}_{L/2},
\end{eqnarray*}
i.e. we arrive at the contradiction. Therefore there exists a value
$\tilde{x}\in[0;L/2]$ such that $\psi(\tilde{x})>\tilde{\psi}_{L/2}$
and $\psi_x(\tilde{x})>0$ i.e. $(\psi(\tilde{x}),\psi_x(\tilde{x}))$ is
$L/2$-collapsing forward point. Then $(\psi(0),\psi_x(0))$ is
$L$-collapsing forward point. So, for $R>\tilde{\psi}'_L$ there are no
points from ${\cal U}_L$ in $S^+_R$. $\blacksquare$
\medskip

\section{Proof of Theorem \ref{singular}.2}\label{Th2}

{\it Proof of Theorem \ref{singular}.2.} Introduce the following
functions:

(a) the function $\tilde{H}^+(\tilde{\psi},\tilde{\psi}',\Omega;t)$
defined as follows:
$\tilde{H}^+(\tilde{\psi},\tilde{\psi}',\Omega;t)=x_0$ if the solution
of Cauchy problem for the equation
\begin{eqnarray*}
\phi_{xx}+\Omega\phi-\phi^3=0
\end{eqnarray*}
with initial data $\phi(t)=\tilde{\psi}$, $\phi_x(t)=\tilde{\psi}'$
collapses at the value $x=x_0$, $x_0>0$. Exact formula for
$\tilde{H}^+(\tilde{\psi},\tilde{\psi}',\Omega;t)$ is
\begin{eqnarray}\label{ExactP3}
\tilde{H}^+(\tilde{\psi},\tilde{\psi}',\Omega;t)=
t+\int_{\tilde{\psi}}^\infty\frac{\sqrt{2}~d\eta}{\sqrt{\eta^4-\tilde{\psi}^4-2{\Omega}(\eta^2-\tilde{\psi}^2)+2(\tilde{\psi}')^2}}
\end{eqnarray}
It follows from  (\ref{ExactP3}) that if
$\tilde{H}^+(\psi_0,\psi'_0,\Omega_0;t)<\infty$ then for $t$ fixed the
function $H^+(\tilde{\psi},\tilde{\psi}',\Omega;t)$ is a continuous
function of the variables $\tilde{\psi},\tilde{\psi}',\Omega$ in some
vicinity of the point $(\psi_0,\psi'_0,\Omega_0)$.\medskip

(b) the function $\tilde{h}^+(\tilde{\psi},\tilde{\psi}';t)$ defined as
follows: $\tilde{h}^+(\tilde{\psi},\tilde{\psi}';t)=x_0$ if the
solution of Cauchy problem for Eq.(\ref{1D_rep}) with initial data
$\psi(t)=\tilde{\psi}$, $\psi_x(t)=\tilde{\psi}'$ collapses at value
$x=x_0$, $x_0>0$. Evidently, if $\psi(x)$ is a solution of
Eq.(\ref{1D_rep}) then
\begin{eqnarray}\label{Transl}
h^+(\psi(0),\psi_x(0))=\tilde{h}^+(\psi(t),\psi_x(t);t)+t
\end{eqnarray}

(c) the two functions
\begin{eqnarray*}
\Omega_1(t)=\min_{x\in[t;L]}(\omega-U(x)),\quad
\Omega_2(t)=\max_{x\in[t;L]}(\omega-U(x))
\end{eqnarray*}
which are continuous functions in some vicinity of the point
$t=L$.\medskip

Also let us denote by $D_\delta(\zeta,\zeta')$ a disc in $\mathbb{R}^2$
with center at $(\zeta,\zeta')$ and radius $\delta$.

It follows from the conditions of Theorem \ref{singular}.2 that the
solution $\psi(x)$ of Eq.(\ref{1D_rep}) with initial data
$\psi(0)=\psi_0$, $\psi_x(0)=\psi_0'$ satisfies one of the conditions
\begin{eqnarray*}
\lim_{x\to L}\psi(x)=+\infty,\quad \mbox{or}\quad \lim_{x\to
L}\psi(x)=-\infty
\end{eqnarray*}
Let the behavior of $\psi(x)$ in vicinity of $x=L$ obey the first of
the two formulas above (the analysis of the second case is similar).
Then there exists $t$ such that $\psi(x)>\sqrt{\Omega_1(t)}$ and
$\psi_x(x)>0$ for $x\in[t;L)$. By virtue of Comparison Lemma, see
\ref{Th1}, one has
\begin{eqnarray}\label{t_star}
\tilde{H}^+(\tilde{\psi},\tilde{\psi}',\Omega_1(t^*);t^*)\leq\tilde{h}^+(\tilde{\psi},\tilde{\psi}';t^*)\leq\tilde{H}^+(\tilde{\psi},\tilde{\psi}',\Omega_2(t^*);t^*)
\end{eqnarray}
for any $t^*\in[t;L)$ and for any $\tilde{\psi},\tilde{\psi}'$ in some
vicinity of the point $(\psi(t^*),\psi'(t^*))$.

Let us describe 3-step algorithm which allows by given $\varepsilon>0$
to find $\delta>0$ such that if $(\psi,\psi')\in D_\delta
(\psi_0,\psi_0')$ then
\begin{eqnarray}\label{ContFinal}
|h^+(\psi,\psi')-h^+(\psi_0,\psi_0')|<\varepsilon.
\end{eqnarray}

1. By given $\varepsilon$ one can find $t^*$ such that the inequality
holds
\begin{eqnarray}\label{App01}
|\tilde{H}^+(\psi(t^*),\psi'(t^*),\Omega_2(t^*);t^*)-\tilde{H}^+(\psi(t^*),\psi'(t^*),\Omega_1(t^*);t^*)|\leq
\varepsilon/2
\end{eqnarray}

2. Since $\tilde{H}^+(\tilde{\psi},\tilde{\psi}',\Omega;t)$ is
continuous there exists $\delta_1>0$ such that when
$(\tilde{\psi},\tilde{\psi}')\in D_{\delta_1}(\psi(t^*),\psi'(t^*))$
the inequalities hold
\begin{eqnarray*}
&&|\tilde{H}^+(\tilde{\psi},\tilde{\psi}',\Omega_2(t^*);t^*)-\tilde{H}^+(\psi(t^*),\psi'(t^*),\Omega_2(t^*);t^*)|\leq
\varepsilon/2\\
&&|\tilde{H}^+(\tilde{\psi},\tilde{\psi}',\Omega_1(t^*);t^*)-\tilde{H}^+(\psi(t^*),\psi'(t^*),\Omega_1(t^*);t^*)|\leq
\varepsilon/2
\end{eqnarray*}
It follows from (\ref{t_star}) and (\ref{App01}) that if
$(\tilde{\psi},\tilde{\psi}')\in D_{\delta_1}(\psi(t^*),\psi'(t^*))$
then
\begin{eqnarray}\label{App02}
|\tilde{h}^+(\tilde{\psi},\tilde{\psi}';t^*)-\tilde{h}^+(\psi(t^*),\psi'(t^*);t^*)|\leq
\varepsilon
\end{eqnarray}

3. The flow defined by Eq.(\ref{1D_rep}) generates a diffeomorphism
$T_{t^*}: \mathbb{R}^2\to\mathbb{R}^2$ which maps a point
$(\psi(t^*),\psi_x(t^*))$ to a point $(\psi(0),\psi_x(0))$ where
$\psi(x)$ is a solution of Eq.(\ref{1D_rep}). Then there exists
$\delta$ such that $D_\delta(\psi_0,\psi_0')\subset
T_{t^*}D_{\delta_1}(\psi(t^*),\psi_x(t^*))$. By means of (\ref{Transl})
and (\ref{App02}) one concludes that for $(\psi,\psi')\in
D_{\delta}(\psi_0,\psi_0')$ the relation (\ref{ContFinal}) holds.
Theorem \ref{singular}.2 is proved. $\blacksquare$

\section{Proof of Theorem \ref{SymDynTheory}.1.}\label{ProofMain}


Before proving Theorem \ref{SymDynTheory}.1 we prove the following
lemma:
\medskip

{\bf Lemma \ref{SymDynTheory}.1.} {\it Let $D$ be an island.

(i) Let $D\supset V_1\supset V_2\supset\ldots$ be an infinite sequence
of nested v-strips such that
\begin{eqnarray}\label{ToZero}
\lim_{n\to\infty}\mu(V_n)=0.
\end{eqnarray}
Then the intersection
\begin{eqnarray*}
V_\infty=\bigcap_{n=1}^\infty V_n
\end{eqnarray*}
is a v-curve.

(ii) Let $D\supset H_1\supset H_2\supset\ldots$ is an infinite sequence
of nested h-strips and
\begin{eqnarray*}
\lim_{n\to\infty}\mu(H_n)=0.
\end{eqnarray*}
Then the intersection
\begin{eqnarray*}
H_\infty=\bigcap_{n=1}^\infty H_n
\end{eqnarray*}
is an h-curve.}

\begin{figure}
\centerline{\includegraphics [scale=0.9]{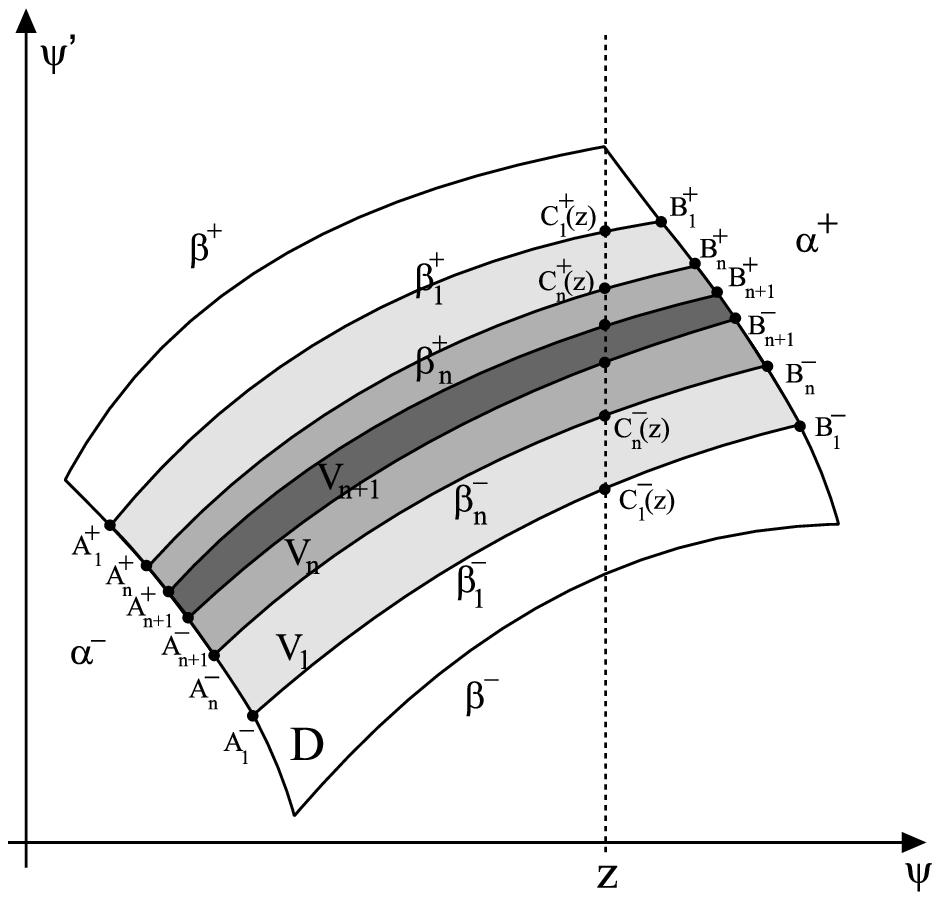}}
\caption{Illustration to the proof of Lemma \ref{SymDynTheory}.1.
}\label{LemmaProof}
\end{figure}

{\it Proof of Lemma \ref{SymDynTheory}.1.} Let us prove the point (i),
the point (ii) can be proved similarly. Denote the v-curves which bound
the strip $V_n$ by $\beta_n^-$ (which lies closer to $\beta^-$) and
$\beta_n^+$ (which lies closer to $\beta^+$). Let the endpoints of
$\beta_n^+$ be $A_n^+$ (situated at $\alpha^-$) and $B_n^+$ (situated
at $\alpha^+$). Let the endpoints of $\beta_n^-$ be $A_n^-$ (situated
at $\alpha^-$) and $B_n^-$ (situated at $\alpha^-$), see
Fig.\ref{LemmaProof}.

{\it First}, we show that  $A_n^-\to A_n^+$ and $B_n^-\to B_n^+$ as
$n\to\infty$. The sequence of points $A_1^-,A_2^-,\ldots$ is situated
on the curve $\alpha^-$ from one side from the point $A_1^+$ and is
``monotonic'' in the sense that for any $n$ the point $A_{n+1}^-$ is
situated on $\alpha^-$ between the points $A_{n}^-$ and $A_1^+$.
Therefore it has a limit point $A_\infty^+\in \alpha^-$. The sequence
of points $A_1^+,A_2^+,\ldots$ is situated on the curve $\alpha^-$ from
one side of the point $A_1^-$ and has similar monotonic property.
Therefore it also has a limit point $A_\infty^-\in \alpha^-$. Suppose
that $A_\infty^+\ne A_\infty^-$. Then, since $\beta_n^\pm$ are graphs
of monotone non-increasing/non-decreasing $\gamma$-Lipschitz functions
and $\alpha^-$ is a graph of monotone non-decreasing/non-increasing
$\gamma$-Lipschitz function, the area of $V_n$ cannot tend to zero as
$n\to\infty$. Therefore $A_\infty^+= A_\infty^-=A_\infty$. In the same
way one can introduce the limit points $B_\infty^-\in \alpha^+$ and
$B_\infty^+\in \alpha^+$ and conclude that $B_\infty^+=
B_\infty^-=B_\infty$.

{\it Second}, let coordinates of $A_\infty$ be $(\psi_A,\psi'_A)$ and
coordinates of $B_\infty$ be $(\psi_B,\psi'_B)$. Consider a real value
$z$ situated between $\psi_A$ and $\psi_B$. Since $A_n^-\to A_n^+$ and
$B_n^-\to B_n^+$ as $n\to\infty$ there exists $n_0$ such that for
$n>n_0$ both $\beta_n^+$ and $\beta_n^-$ intersect the line $\psi=z$.
Denote the points of intersections of $\beta_n^+$ and $\beta_n^-$ with
the line $\psi=z$ correspondingly $C_n^+(z)$ and $C_n^-(z)$. Evidently,
both the sequences $\{C_n^+(z)\}$ and $\{C_n^-(z)\}$ have limits as
$n\to\infty$. Denote these limits $C^+(z)$ and $C^-(z)$
correspondingly.

Assume that at some $z=z^*$ one has $|C^+(z)-C^-(z)|=\delta_C> 0$. The
relation $C^+(z)\ne C^-(z)$ cannot hold in some vicinity of the point
$z=z^*$, otherwise there exists a set of nonzero measure which belongs
to all the nested strips $V_n$ and $\mu(V_n)$ does not tend to zero as
$n\to \infty$. Therefore for any $\varepsilon$ there exist a value
$z_1$, such that $|z_1-z^*|<\varepsilon$ and $C^+(z_1)= C^-(z_1)$. This
means that (al least) one of the ratios
\begin{eqnarray*}
\frac{|C^+(z_1)-C^+(z^*)|}{|z_1-z^*|},\quad
\frac{|C^-(z_1)-C^-(z^*)|}{|z_1-z^*|}
\end{eqnarray*}
is greater than $\delta_C/2\varepsilon$. Since $\varepsilon$ can be
taken arbitrarily small, this contradicts the condition that
$\beta_n^\pm$ are graphs of monotone non-increasing/non-decreasing
$\gamma$-Lipschitz functions. This implies that $C^+(z)= C^-(z)\equiv
C(z)$ for all $z$, $\psi_A<z<\psi_B$.

{\it Third}, each of the curves $\beta_n^-$ is a graph of a monotone
non-increasing/non-decreasing $\gamma$-Lipschitz function. Passing to
the limit $n\to\infty$ we obtain a curve consisting of the points
$C(z)$, $\psi_A<z<\psi_B$. This curve is also a graph of monotone
non-increasing/non-decreasing $\gamma$-Lipschitz function with the same
$\gamma$, (see \cite{Wiggins}, Sect 4.3), i.e. v-curve.$\blacksquare$
\medskip

{\it Proof of Theorem \ref{SymDynTheory}.1.} Evidently, for each ${\bf
p}\in {\cal P}$ the image ${\bf s}=\Sigma{\bf p}\in \Omega^N$ is
defined uniquely. Let us prove that for each ${\bf s}\in \Omega^N$
there exists unique ${\bf p}\in {\cal P}$ such that ${\bf s}=\Sigma{\bf
p}$. Consider a sequence ${\bf s}=\{\ldots,i_{-1},i_0,i_1,\ldots\}$,
$i_k\in\{1,\ldots,N\}$. Let us find the location of the points $p\in
D_{i_0}$ such that $T^{-1}p\in D_{i_{-1}}$, $T^{-2}p\in D_{i_{-2}}$
etc. It is easy to check that:

- the points $p\in D_{i_0}$ such that $T^{-1}p\in D_{i_{-1}}$ are
situated in the set $V_{i_{-1}i_0}=TD_{i_{-1}}\cap D_{i_0}$. Due to the
condition (i) of theorem, $V_{i_{-1}i_0}$ is a v-strip. Moreover,
$V_{ji_0}$ and $V_{ki_0}$ have no common points if $j\ne k$.

- the points $q\in D_{i_{-1}}$ such that $T^{-1}q\in D_{i_{-2}}$ are
situated in the set $V_{i_{-2}i_{-1}}=TD_{i_{-2}}\cap D_{i_{-1}}$ which
is a v-strip. The points $p\in D_{i_0}$ such that $T^{-1}p=q\in
D_{i_{-1}}$, $T^{-2}p=T^{-1}q\in D_{i_{-2}}$ are situated in the set
$V_{i_{-2}i_{-1}i_0}=TV_{i_{-2}i_{-1}}\cap D_{i_0}$ which is also
v-strip. Evidently, $V_{i_{-2}i_{-1}i_0}\subset V_{i_{-1}i_0}$. Also,
$V_{ji_{-1}i_0}$ and $V_{ki_{-1}i_0}$ have no common points if $j\ne
k$.

Continuing the process, we have nested sequence of v-strips
\begin{eqnarray*}
\ldots V_{i_{-(n+1)}i_{-n}\ldots i_{-2}i_{-1}i_0}\subset
V_{i_{-n}\ldots i_{-2}i_{-1}i_0}\subset\ldots \subset
V_{i_{-2}i_{-1}i_0}\subset V_{i_{-1}i_0}\subset D_{i_0}
\end{eqnarray*}
such that $V_{i_{-(n+1)}i_{-n}\ldots
i_{-2}i_{-1}i_0}=TV_{i_{-(n+1)}i_{-n}\ldots i_{-2}i_{-1}}\cap D_{i_0}$.
Since
\begin{eqnarray*}
\mu(V_{i_{-n}\ldots i_{-2}i_{-1}i_0})\leq \mu(\Delta^+_n)
\end{eqnarray*}
the area of v-strip $V_{i_{-n}\ldots i_{-2}i_{-1}i_0}$ tends to zero as
$n\to\infty$. According to Lemma \ref{SymDynTheory}.1 the intersection
of these nested strips, $V_\infty$, exists and is a v-curve.

In the same manner the nested sequence of h-strips can be constructed,
\begin{eqnarray*}
\ldots H_{i_0i_1\ldots i_{n}i_{n+1}}\subset H_{i_0i_1\ldots
i_{n}}\subset\ldots \subset H_{i_0i_1i_2}\subset H_{i_0i_1}\subset
D_{i_0}
\end{eqnarray*}
where $H_{i_0i_1\ldots i_{n}i_{n+1}}=T^{-1}H_{i_0i_1\ldots i_{n}}\cap
D_{i_0}$. The area of the strip $H_{i_0i_1\ldots i_{n}}$ tends to zero
as $n\to\infty$ so according to Lemma \ref{SymDynTheory}.1 the
intersection of these nested strips, $H_\infty$, exists and is an
h-curve.

The orbit ${\bf p}\in {\cal P}$ corresponding to  bi-infinite sequence
${\bf s}=\{\ldots,i_{-1},i_0,i_1,\ldots\}$ is generated by $T$- and
$T^{-1}$-iterations of the intersection $H_\infty\cap V_\infty$ which
according to the definition of h- and v-curves consists of one point.
Therefore ${\bf p}$  exists and is unique.

The continuity of $\Sigma$ and $\Sigma^{-1}$ follows from the following
observations:\medskip

- since $T$ is continuous, if ${\bf
p}^{(1)}=\{\ldots,p_{-1}^{(1)},p_{0}^{(1)},p_{1}^{(1)},\ldots\}$ and
${\bf p}^{(2)}=\{\ldots,p_{-1}^{(2)},p_{0}^{(2)},p_{1}^{(2)},\ldots\}$
are close enough in ${\cal P}$ (i.e. the points $p_{0}^{(1)}$ and
$p_{0}^{(2)}$ are close in $\mathbb{R}^2$), then  their $\Sigma$-images
share the same central block $|i|<k$ for some $k$. Therefore they are
also close in $\Omega^N$-topology;\medskip

- if ${\bf s}^{(1)}=\Sigma{\bf p}^{(1)}$ and ${\bf s}^{(2)}=\Sigma{\bf
p}^{(2)}$ share the same central block $|i|<k$ for some $k$, the points
$p_{0}^{(1)}$ and $p_{0}^{(2)}$ are situated in the curvilinear
quadrangle $V_{i_{-k}\ldots i_{-2}i_{-1}i_0}\cap H_{i_0i_1\ldots
i_{k}}$, so ${\bf p}^{(1)}$ and ${\bf p}^{(2)}$ are close in ${\cal
P}$-topology. $\blacksquare$

\section{Proof of Theorem \ref{CodingGen}.1}\label{ProofHyp2}

Let ${\bf e}_{1,2}$ be unit vectors (\ref{UnitV}). Denfine the
following cones
\begin{eqnarray*}
&&\mathbb{R}^2_{++}=\{{\bf a}|~ {\bf a}=x{\bf e}_1+y{\bf
e}_2,~x>0,y>0\},\\
&&\overline{\mathbb{R}}^2_{++}=\{{\bf a}|~ {\bf a}=x{\bf e}_1+y{\bf
e}_2,~x\geq0,y\geq0\},\\
&&\mathbb{R}^2_{-+}=\{{\bf a}|~ {\bf a}=x{\bf e}_1+y{\bf
e}_2,~x>0,y<0\},\\
&&\overline{\mathbb{R}}^2_{-+}=\{{\bf a}|~ {\bf a}=x{\bf e}_1+y{\bf
e}_2,~x\geq0,y\leq0\},\\
&&\mathbb{R}^2_{+-}=\{{\bf a}|~ {\bf a}=x{\bf e}_1+y{\bf
e}_2,~x>0,y<0\},\\
&&\overline{\mathbb{R}}^2_{+-}=\{{\bf a}|~ {\bf a}=x{\bf e}_1+y{\bf
e}_2,~x\geq0,y\leq0\},\\
&&\mathbb{R}^2_{--}=\{{\bf a}|~ {\bf a}=x{\bf e}_1+y{\bf
e}_2,~x<0,y<0\},\\
&&\overline{\mathbb{R}}^2_{--}=\{{\bf a}|~ {\bf a}=x{\bf e}_1+y{\bf
e}_2,~x\leq0,y\leq0\}.
\end{eqnarray*}

{\bf Lemma \ref{CodingGen}.1.} {\it Let $S\subset {\mathbb R}^2$ be a
compact connected set, $T$ be a diffeomorphism defined on $S$,  the
operator $DT_{p}$ be nondegenerate for all $p\in S$ and ${\bf
e}_1$,${\bf e}_2$ be the unit vectors defined by (\ref{UnitV}).

I. Let for all $p\in S$ the relations $g_1(p)>0$ and $g_2(p)>0$ hold.
Then

a. For all $p\in S$ one and only one of the following alternative
conditions holds
\begin{eqnarray*}
&{\rm [A1]}& ~DT_{p} [\overline{\mathbb{R}}^2_{++}]\subset
\mathbb{R}^2_{++};\quad {\rm [A2]}~ DT_{ p}
[\overline{\mathbb{R}}^2_{++}]\subset \mathbb{R}^2_{--};\\
&{\rm [A3]}& ~DT_{ p} [\overline{\mathbb{R}}^2_{++}]\subset
\mathbb{R}^2_{+-};\quad {\rm [A4]}~ DT_{ p}
[\overline{\mathbb{R}}^2_{++}]\subset \mathbb{R}^2_{-+}
\end{eqnarray*}

b. there exists $\gamma>0$ such that for any two points
$p_1=(\psi_1,\psi_1')\in S$ and $p_2=(\psi_2,\psi_2')\in S$,
$\psi_1<\psi_2$ and $\psi'_1<\psi'_2$ the images
$Tp_1=q_1=(\phi_1,\phi_1')$, $Tp_2=q_2=(\phi_2,\phi_2')$ are such that
\begin{eqnarray*}
\mbox{in the case [A1]}:\quad 0<\phi_2-\phi_1<\gamma(\phi'_2-\phi'_1)\\
\mbox{in the case [A2]}:\quad 0<\phi_1-\phi_2<\gamma(\phi'_1-\phi'_2)\\
\mbox{in the case [A3]}:\quad 0<\phi_1-\phi_2<\gamma(\phi'_2-\phi'_1)\\
\mbox{in the case [A4]}:\quad 0<\phi_2-\phi_1<\gamma(\phi'_1-\phi'_2)
\end{eqnarray*}

II. Let for all $p\in S$ the relations $g_1(p)<0$ and $g_2(p)<0$ hold.
Then

a. For all $p\in S$ one and only one of the following alternative
conditions holds
\begin{eqnarray*}
&{\rm [B1]}& ~DT_{ p} [\overline{\mathbb{R}}^2_{-+}]\subset
\mathbb{R}^2_{++};\quad {\rm [B2]}~ DT_{ p}
[\overline{\mathbb{R}}^2_{-+}]\subset \mathbb{R}^2_{--};\\
&{\rm [B3]}& ~DT_{ p} [\overline{\mathbb{R}}^2_{-+}]\subset
\mathbb{R}^2_{+-};\quad {\rm [B4]}~ DT_{ p}
[\overline{\mathbb{R}}^2_{-+}]\subset \mathbb{R}^2_{-+}
\end{eqnarray*}

b. There exists $\gamma>0$ such that for any two points
$p_1=(\psi_1,\psi_1')\in S$ and $p_2=(\psi_2,\psi_2')\in S$,
$\psi_1<\psi_2$ and $\psi'_1>\psi'_2$ the images
$Tp_1=q_1=(\phi_1,\phi_1')$, $Tp_2=q_2=(\phi_2,\phi_2')$ are such that
\begin{eqnarray*}
\mbox{in the case [B1]}:\quad 0<\phi_2-\phi_1<\gamma(\phi'_2-\phi'_1)\\
\mbox{in the case [B2]}:\quad 0<\phi_1-\phi_2<\gamma(\phi'_1-\phi'_2)\\
\mbox{in the case [B3]}:\quad 0<\phi_1-\phi_2<\gamma(\phi'_2-\phi'_1)\\
\mbox{in the case [B4]}:\quad 0<\phi_2-\phi_1<\gamma(\phi'_1-\phi'_2)
\end{eqnarray*}
}

{\it Proof of Lemma \ref{CodingGen}.1.} Let us prove the point Ia, the
point IIa can be proved similarly. Evidently, the relations $g_1(p)>0$
and $g_2(p)>0$ mean that both the vectors $DT_{p}{\bf e}_1$,
$DT_{p}{\bf e}_2$ are situated in the same quadrant,
$\mathbb{R}^2_{++}$, $\mathbb{R}^2_{-+}$, $\mathbb{R}^2_{+-}$ or
$\mathbb{R}^2_{--}$ for any $p\in S$. Let for two points $p_1$ and
$p_2$ these quadrants are different. Connect these points by continuous
curve $\zeta\subset S$. Since $DT_{p}{\bf e}_1$, $DT_{p}{\bf e}_2$
depend continuously on $p$ there exist a point $p^*\in \zeta$, such
that one of $(DT_{p}{\bf e}_i,{\bf e}_j)$, $i=1,2$, $j=1,2$ vanishes
therefore $g_1(p^*)=0$ or $g_2(p^*)=0$. This implies that the point I
is valid.

Let us prove the point Ib. Assume that $g_1(p)>0$ and $g_2(p)>0$ hold
and the situation [A1] takes place, the situations [A2]-[A4] can be
treated similarly. It follows from the condition [A1] and compactness
of $S$ that there exists a supremum
\begin{eqnarray*}
\tilde \gamma=\sup \frac{\xi_2}{\xi_1},\quad  \left(
\begin{array}{c}
\xi_1\\\xi_2
\end{array}\right)=DT_pz,\quad z\in \overline{\mathbb{R}}^2_{++},\quad p\in
S.
\end{eqnarray*}
 Let $\psi_2>\psi_1$ and
$\psi_2'>\psi_1'$ and $p_1=(\psi_1,\psi_1')$, $p_2=(\psi_2,\psi_2')$,
$q_1=Tp_1=(\phi_1,\phi_1')$, $q_2=Tp_2=(\phi_2,\phi_2')$. Then
\begin{eqnarray*}
\left(
\begin{array}{c}
\phi_2-\phi_1\\\phi'_2-\phi'_1
\end{array} \right)=DT_{{p}_1} \left(
\begin{array}{c}
\psi_2-\psi_1\\\psi'_2-\psi'_1
\end{array}\right) +{r}
\end{eqnarray*}
where $\|{r}\|/\|{p}_2-{p}_1\|\to 0$ when $\|{ p}_2-{p}_1\|\to 0$. This
implies that for close enough  ${p}_1$ and ${p}_2$ one can choose
$\gamma>\tilde \gamma$ such that for corresponding ${q}_1$ and ${q}_2$
\begin{eqnarray}
0<\phi'_2-\phi'_1<\gamma(\phi_2-\phi_1)\label{Transit}
\end{eqnarray}
This ordering  is transitive: from the relation (\ref{Transit}) and the
relation $0<\phi'_3-\phi'_2<\gamma(\phi_3-\phi_2)$ it follows that
$0<\phi'_3-\phi'_1<\gamma(\phi_3-\phi_1)$. Therefore we can omit the
words ``for close enough'' above and state that (\ref{Transit}) holds
for any ${p}_{1,2}$ such that $\psi_2>\psi_1$ and $\psi_2'>\psi_1'$.
So, the point Ib under the assumption [A1] is proved. In the same
manner the point Ib can be proved for other three cases, [A2]-[A4]. The
proof of the point IIb consists in considering in the same manner the
situations [B1]-[B4]. $\blacksquare$
\medskip

{\it Proof of Theorem \ref{CodingGen}.1.}  Since Eq.(\ref{1D_rep}) is
invariant with respect to $x$-inversion, the strip ${\cal U}^+_\pi$ is
symmetric with respect to the origin. If a point
$(\tilde{\psi};\tilde{\psi}')$ is situated on one edge, $\alpha^+$, of
the strip ${\cal U}^+_\pi$ the solution $\psi(x)$ of Cauchy problem
with initial data $(\tilde{\psi};\tilde{\psi}')$ obeys the condition
\begin{eqnarray}
\lim_{x\to\pi}\psi(x)=+\infty\label{+inf}
\end{eqnarray}
whereas for initial data on another edge, $\alpha^-$, of ${\cal
U}^+_\pi$ the corresponding condition is
\begin{eqnarray}
\lim_{x\to\pi}\psi(x)=-\infty\label{-inf}
\end{eqnarray}
The set ${\cal U}^-_\pi$ is also infinite curvilinear strip related to
${\cal U}^+_\pi$ by the symmetry with respect to the axis $\psi$.

Let $V$ be v-strip situated in an island $D_i$ between two v-curves
$\tilde\beta^+_i$ and $\tilde\beta^-_i$. $V$ is curvilinear quadrangle
bounded by $\tilde\beta^+_i$ and $\tilde\beta^-_i$ and two more bounds
lying on $\alpha^+$ and $\alpha^-$. Taking into account (\ref{+inf})
and (\ref{-inf}) one concludes that $TV$ is an infinite curvilinear
strip stretching along ${\cal U}^-_\pi$ having the edges
$T\tilde\beta^+_i$ and $T\tilde\beta^-_i$. This means that $TV$ crosses
all $\alpha_j^\pm$, $j=1,\ldots,N$ at least once and pass through all
the sets $D_j$, $j=1,\ldots,N$.

Let a pair $(i,j)$ be fixed. Assume that the curves $\beta^\pm_i$ are
graphs of monotone non-decreasing functions. Then $\tilde\beta^+_i$ are
also graphs of monotone non-decreasing functions. Let for all $p\in
T^{-1}D_j\cap D_i$ the conditions $g_1(p)>0$ and $g_2(p)>0$ hold. It
follows from Lemma \ref{CodingGen}.1 that for all the points $p\in
T^{-1}D_j\cap D_i$ only one of the conditions [A1]-[A4] hold. This
means that the images $T\tilde\beta_i^+\cap D_j$ and
$T\tilde\beta_i^-\cap D_j$ consist of one connected component. In fact,
if $T\tilde\beta_i^+$ crosses $\alpha_j^+$ or $\alpha_j^-$ twice one
can choose two pairs of points $p_{1,2}\in\tilde\beta_i^+$,
$p_{3,4}\in\tilde\beta_i^+$, 
\begin{eqnarray*}
&&p_1=(\psi_1,\psi'_1),\quad p_2=(\psi_2,\psi'_2),\quad
\psi_1<\psi_2,\quad \psi_1'<\psi_2'\\
&&p_3=(\psi_3,\psi'_3),\quad p_4=(\psi_4,\psi'_4),\quad
\psi_3<\psi_4,\quad \psi_3'<\psi_4'
\end{eqnarray*}
such that their images $q_k=Tp_k\in D_j$,
\begin{eqnarray*}
q_1=(\phi_1,\phi'_1),\quad q_2=(\phi_2,\phi'_2),\quad
q_3=(\phi_3,\phi'_3),\quad q_4=(\phi_4,\phi'_4)
\end{eqnarray*}
are mismatched in the sense that at least one of the products
\begin{eqnarray*}
(\phi_2-\phi_1)(\phi_4-\phi_3)\quad\mbox{or}\quad
(\phi_2'-\phi_1')(\phi_4'-\phi_3')
\end{eqnarray*}
is negative. By Lemma \ref{CodingGen}.1 the images
$T\tilde\beta_i^\pm\cap D_j$ are  graphs of non-decreasing or
non-increasing $\gamma$-Lipschitz functions. Since $D_j$ is an island,
the boundaries $\beta_j^\pm$ are graphs of monotonic functions and by
geometric reasons the  monotonicity properties (non-increasing or
non-decreasing) of $\beta_j^\pm$ and $T\tilde\beta_i^\pm\cap D_j$ are
the same. Therefore $T\tilde\beta_i^\pm\cap D_j$ are v-curves. They
bounded the set $TV\cap D_j$, therefore $TV\cap D_j$ is a v-strip.

If the curves $\beta^\pm_i$ are graphs of monotone non-increasing
functions and for all $p\in T^{-1}D_j\cap D_i$ the conditions
$g_1(p)<0$ and $g_2(p)<0$ hold the proof repeats the reasoning given
above making use of the conditions [B1]-[B4]. Theorem \ref{CodingGen}.1
is proved. $\blacksquare$

\end{document}